\documentclass[colorlinks=true, linkcolor=blue, citecolor=blue, urlcolor=blue]{aa}  
\usepackage{orcidlink}
\usepackage{txfonts}
\usepackage{graphicx}	
\usepackage{amsmath}	
\usepackage{amssymb}	
\usepackage{xcolor}     
\usepackage{soul}       
\usepackage{array}
\usepackage{makecell}
\usepackage{enumitem}
\usepackage{arydshln}
\usepackage{pdflscape}
\usepackage{multirow}
\usepackage{etoolbox}
\usepackage{placeins}

\def\zlmean{$z_{\rm d, mean} = 0.645$\xspace}
\def\zsmean{$z_{\rm s, mean} = 2.410$\xspace}

\newcommand{\sref}[1]{Section~\ref{#1}}

\makeatletter
\newcommand{\mycustombox}[1]{%
  \fbox{%
    \begin{minipage}{\linewidth-2\fboxsep-2\fboxrule}
      \vspace{0.15cm} 
      
      \raggedleft 
      \textbf{Prepared for Submission} --- \textit{Physical Review D}\\
      Draft version: \today
      
      \vspace{0.15cm}
    \end{minipage}%
  }%
}
\patchcmd{\aa@maketitle}{\fbox}{\mycustombox}{}{}
\makeatother

\makeatletter
\newcommand{\customcite}[2]{\hyper@@link[cite]{}{cite.#1}{#2}}
\renewcommand*\aa@textidlineempty{\aa@headings}
\renewcommand*\aa@headfont{\small\mathversion{normal}}
\makeatother

\begin{document} 

\title{\textit{HST} imaging, pipeline modeling, and time-delay predictions of 2 triply-imaged and 15 quadruply-imaged lensed quasars}
\author{William~Sheu\orcidlink{0000-0003-1889-0227}\inst{\ref{af:ucla}}\fnmsep\thanks{Corresponding author: wsheu@astro.ucla.edu}
\and Tommaso~Treu\orcidlink{0000-0002-8460-0390}\inst{\ref{af:ucla}}
\and Adriano~Agnello\orcidlink{0000-0001-9775-0331}\inst{\ref{af:hartree}}
\and Timo~Anguita\orcidlink{0000-0003-0930-5815}\inst{\ref{af:bello}}
\and Simon~Birrer\orcidlink{0000-0003-3195-5507}\inst{\ref{af:stoney}}
\and Daniel~Gilman\orcidlink{0000-0002-5116-7287}\inst{\ref{af:uchicago}}
\and Xiaosheng~Huang\orcidlink{0000-0001-8156-0330}\inst{\ref{af:usf}, \ref{af:lbnl}}
\and Richard~G.~McMahon\orcidlink{0000-0001-8447-8869}\inst{\ref{af:ucambridge1}, \ref{af:ucambridge2}}
\and Nicholas~D.~Morgan\orcidlink{0000-0001-7779-9883}\inst{\ref{af:staples}}
\and Veronica~Motta\orcidlink{0000-0003-4446-7465}\inst{\ref{af:valpara}}
\and Anna~Nierenberg\orcidlink{0000-0001-6809-2536}\inst{\ref{af:merced}}
\and Kenneth~C.~Wong\orcidlink{0000-0002-8459-7793}\inst{\ref{af:tokyo}}
\and Ioana~Zelko\orcidlink{0000-0002-7588-976X}\inst{\ref{af:utoronto}}
}
\institute{Department of Physics and Astronomy, University of California, Los Angeles, CA 90095, USA \label{af:ucla}
\and STFC Hartree Centre, Sci-Tech Daresbury, Keckwick Lane, Daresbury, Warrington, WA4 4AD, United Kingdom \label{af:hartree}
\and Instituto de Astrof\'{i}sica, Facultad de Ciencias Exactas, Universidad Andres Bello, Fern\'{a}ndez Concha 700, 7591538, Las Condes, Santiago, Chile \label{af:bello}
\and Department of Physics and Astronomy, Stony Brook University, Stony Brook, NY 11794, USA \label{af:stoney}
\and Department of Astronomy and Astrophysics, University of Chicago, Chicago, IL 60637, USA \label{af:uchicago}
\and Department of Physics \& Astronomy, University of San Francisco, San Francisco, CA 94117, USA \label{af:usf}
\and Physics Division, Lawrence Berkeley National Laboratory, 1 Cyclotron Road, Berkeley, CA 94720, USA \label{af:lbnl}
\and Institute of Astronomy, University of Cambridge, Madingley Road, Cambridge CB3 0HA, UK \label{af:ucambridge1}
\and Kavli  Institute for Cosmology, University of Cambridge, Madingley Road, Cambridge CB3 0HA, UK \label{af:ucambridge2}
\and Staples High School, 70 North Avenue, Westport, CT 06880, USA \label{af:staples}
\and Instituto de F\'{\i}sica y Astronom\'{\i}a, Universidad de Valpara\'{\i}so, Avda. Gran Breta\~na 1111, Valpara\'{\i}so, Chile. \label{af:valpara}
\and University of California, Merced, 5200 N Lake Road, Merced, CA 95341, USA \label{af:merced}
\and Research Center for the Early Universe, Graduate School of Science, The University of Tokyo, 7-3-1 Hongo, Bunkyo-ku, Tokyo 113-0033, Japan \label{af:tokyo}
\and University of Toronto, 27 King's College Cir, Toronto, ON M5S 1A1, Canada \label{af:utoronto}
}
\date{\hspace*{\fill} \today \hspace*{\fill}}
\abstract
 {The Hubble tension remains a significant challenge in modern cosmology, exhibiting a discrepancy between early-Universe cosmic microwave background measurements and local distance ladder observations. Strong lensing time-delay cosmography provides an independent, geometric probe of $H_0$ that can help resolve this discrepancy. Although hundreds of lensed quasars have been discovered, only a handful have been analyzed due to the resource-intensive follow-up required to measure precise time delays and break degeneracies. We present uniform gravitational lens modeling of 17 recently discovered lensed quasar systems (2 triply-imaged and 15 quadruply-imaged) to identify and prioritize the most promising candidates for future cosmological study. Using high-resolution near-infrared \textit{Hubble Space Telescope} WFC3/IR F160W imaging (PID: 17916, PI: T. Treu), we perform uniform pipeline modeling with \textsc{Lenstronomy}. We constrain the mass and light profiles of the deflector galaxies, and assuming a fiducial cosmology, we predict their Fermat potential differences and expected time delays. Our pipeline successfully yields models and time-delay predictions for all 17 systems. Assuming ideal monitoring conditions, we estimate the total contribution from time-delay and Fermat potential modeling errors to the time-delay distance. From this, we classify the systems by estimated time-delay distance uncertainties: six ``excellent'' ($\leq 3\%$), five ``good'' ($3\%$--$7\%$), three ``suitable'' ($7\%$--$12\%$), and three ``impractical'' ($>12\%$). We recommend prioritizing follow-up campaigns on the 11 ``excellent'' and ``good'' systems, which have the potential to deliver high-precision, independent constraints on $H_0$ to help resolve the Hubble tension.}
\keywords{Gravitational lensing: strong -- Cosmology: cosmological parameters -- Cosmology: distance scale}

\titlerunning{\textit{HST} imaging and modeling of 17 lensed quasars}
\authorrunning{W. Sheu et al.}
\maketitle
\nolinenumbers

%
\section{Introduction} \label{sec:introduction}
Measurements of the Hubble-Lema\^{\i}tre constant ($H_0$), which determines the expansion rate and age scale of the Universe, exhibit a statistically significant tension (the ``Hubble tension'') between early-Universe probes assuming a flat $\Lambda\mathrm{CDM}$ model and late-Universe local measurements. Specifically, measurements of the cosmic microwave background from the \textit{Planck} satellite yield a low value of $H_0 \approx 67.4 \pm 0.5$ km s$^{-1}$ Mpc$^{-1}$ \citep{Planck2020}, while local distance ladder measurements using Cepheid variables and Type Ia supernovae yield a higher value of $H_0 \approx 73.04 \pm 1.04$ km s$^{-1}$ Mpc$^{-1}$ \citep{Riess2022}. Resolving this discrepancy is one of the most pressing challenges in modern cosmology, prompting the search for independent, systematic-free cosmological probes \citep{valentino2021}. 

Strongly lensed quasars offer a powerful, independent alternative through time-delay cosmography, first proposed by \citet{Refsdal1964}. In a strongly lensed system, the variability of a background quasar travels along different paths through the gravitational potential of the foreground deflector galaxy, resulting in multiple images on the sky with observable relative time delays ($\Delta t$).  These delays are directly related to the Fermat potential differences between images ($\Delta \Phi$, which are measured through lens modeling) as:
\begin{equation} \label{eq:td}
     \Delta t = \frac{D_{\Delta t}}{c}\Delta \Phi.
\end{equation}
The time-delay distance $D_{\Delta t}$ combines the angular diameter distances to the lens deflector ($D_{\rm d}$), to the source ($D_{\rm s}$), and between them ($D_{\rm ds}$): 
\begin{equation} \label{eq:tdd}
    D_{\Delta t} \equiv (1 + z_{\rm d}) \frac{D_{\rm d} D_{\rm s}}{D_{\rm ds}}.
\end{equation}
This combination of angular diameter distances is inversely proportional to $H_0$, making it an invaluable tool for constraining cosmological models \citep{TDCOSMO2025}.

Although time-delay cosmography is a clean geometric probe, converting observed time delays into precise cosmological constraints requires an accurate model of the gravitational potential of the deflector galaxy. A primary source of systematic uncertainty is the mass-sheet degeneracy \citep[MSD;][]{Falco:1985}, a theoretical transform that scales the mass profile of the lens and adds a constant density sheet. This transform leaves the lensed image positions and flux ratios unchanged but scales the predicted time delays for a given cosmology, thereby directly affecting the inferred value of $H_0$.  To break this degeneracy to the degree necessary for precise cosmography analysis, non-lensing data are needed, such as the stellar kinematics of the deflector.  This information, in combination with a robust lens model, can be used to constrain a dynamical model used to break the MSD \citep[e.g.,][]{Shajib:2023,sheu2026}.  To model these complex systems, state-of-the-art software tools like \textsc{Lenstronomy} \citep{birrer2018, lenstronomyII} are utilized within the TDCOSMO framework to automate uniform lens modeling and quantify systematic differences between modeling methodologies \citep[e.g.,][]{Ding2021, Shajib_2022, Schmidt_2022, Ertl_2023, brady2026}.  These models not only constrain the dynamical model, but they also inform future studies on which systems are capable of achieving this cosmology-level precision, allowing for better allocation of telescope resources, such as for IFU or time-delay observations.

Over the past decade, wide-field photometric and astrometric surveys have revolutionized the discovery of strongly lensed quasars. Blind and targeted searches utilizing data from \textit{Gaia} \citep{gaia2016, Delchambre_2019, Stern_2021, Lemon_2022}, Pan-STARRS \citep{panstars, Lemon_2022}, DES \citep{agnello2018, Treu2018,Agnello_2019}, and the DESI Legacy Imaging Surveys \citep{desils, Dawes_2023, He_2023, Storfer2024, Sheu_2024} have expanded the sample of known lensed quasars by orders of magnitude. However, discoveries from survey imaging are often limited by atmospheric seeing, which blends the light of the deflector galaxy with the bright, point-like quasar images. This blending obscures the lensed host galaxy arcs and prevents precise astrometric positioning of the images. High-resolution imaging is therefore required to cleanly decompose the lens and source light. By observing with the \textit{Hubble Space Telescope} (\textit{HST}) WFC3/IR F160W filter, we probe the rest-frame optical emission of the source host galaxy, which generally exhibits a higher surface brightness than shorter wavelengths. This allows for the high signal-to-noise detection of lensed arcs and a more informed lens model.  

In this paper, we present uniform models of a sample of 17 recently discovered lensed quasars (2 triply-imaged and 15 quadruply-imaged systems) using high-resolution IR \textit{HST} F160W imaging.  Our sample represents a diverse set of lensing configurations, including rare triply-imaged systems such as the ``straddled saddle'' J0316-4106 and the ``naked-cusp'' J0457-7820, as well as the low-luminosity Type-2 AGN lensed system DELVEJ1258-0319 \citep{Schechter_2025}. Using \textsc{Lenstronomy}, we build a robust pipeline-based modeling workflow that models the lens galaxies with joint-constrained bulge-halo S\'ersic light profiles and elliptical power-law mass profiles with external shear, while representing the source host galaxy light with a Sérsic profile with a shapelets basis.  We present predicted Fermat potential differences, time delays, and estimated time-delay distance uncertainties (assuming ideal observing conditions) for the entire sample, providing a valuable foundation for future time-delay monitoring campaigns, spatially-resolved spectroscopy, or a wide variety of additional auxiliary lensing science \citep[e.g.,][]{Brada_2004, Keeley_2025, odonnell2026, urcelay2026}.


This paper is structured as follows. In \sref{sec:hstimaging}, we describe the \textit{HST} observations and our subsampling procedure. In \sref{sec:quasarsample}, we detail the lensed quasar sample, providing individual descriptions for each system. In \sref{sec:methodology}, we outline our uniform modeling pipeline. In \sref{sec:results}, we present our models findings, including the primary lens mass, light, magnification, time delays, and time-delay distance prediction  results. Finally, we conclude and summarize our findings in \sref{sec:conclusion}.  Our full model figures, posteriors, and supplemental information are found in Appendix~\ref{app:A}.

\section{\textit{HST} F160W imaging} \label{sec:hstimaging}
From April 2025 to May 2026, \textit{HST} observed our sample of 17 lensed quasars in the F160W WFC3-IR filter for 2497 seconds across four dithers, as part of the PID:~17916 \textit{HST}-GO program (PI: T. Treu). 

Observations of strongly lensed quasar systems in the near-infrared, particularly utilizing \textit{HST}'s WFC3/IR F160W filter, present a well-documented trade-off between foreground lens contamination and enhanced morphological detail in the background source. Massive elliptical lens galaxies typically reside at intermediate redshifts ($z_{\rm d} \sim 0.2 - 0.6$), meaning their older, redder stellar populations are more luminous in the observed near-infrared than in the optical. While this results in substantial blending and contamination of the lensed arcs, the application of robust lens modeling techniques allows for the accurate characterization and subtraction of this foreground emission. Once the lens light is properly modeled, the F160W band becomes a critical tool for analyzing the extended structure of the source quasar host galaxy. While the highly structured, clumpy morphologies typical of the rest-frame ultraviolet are desirable for providing tight constraints on the lens model, capturing these features in bluer optical bands is often limited by the intrinsic faintness of high-redshift ($z_s > 1$) sources and sometimes by dust effects.  The near-infrared window directly probes the rest-frame optical emission. By capturing the bulk of the host galaxy's established stellar mass, the F160W band yields the signal-to-noise ratio necessary for precision modeling that optical data alone frequently fail to provide.

\subsection{Subsampling}
Utilizing the four dithered exposures, we are able to drizzle each exposure from a native pixel scale of $0\farcs13$/pixel to $0\farcs0975$/pixel.  While it is mathematically possible to subsample up to half the native pixel scale assuming an optimal drizzling pattern, it would also introduce correlated ``ripple-like'' noise near bright objects; this is a result of sampling near the edge of the allowed Nyquist regime.  The extent to which this affects an observation can vary greatly depending on the system observed.  As the goal of this project is to build a pipeline to model all 17 lensed quasar systems, we opt to use a more conservative final pixel size of 75\% of the original.  This mitigates the correlated noise effect while also providing a substantial improvement to the data sampling.  Due to a failure of the guide star acquisition, we only have three dithers of DESIJ2321-0330.  Despite this, we are still well within the regime of Nyquist sampling and we find no evidence of the correlated noise in our models, albeit with a slightly lower signal-to-noise (S/N).

\section{Lensed quasar sample} \label{sec:quasarsample}
\begin{figure*}
\begin{center}
 \includegraphics[width=1\linewidth]{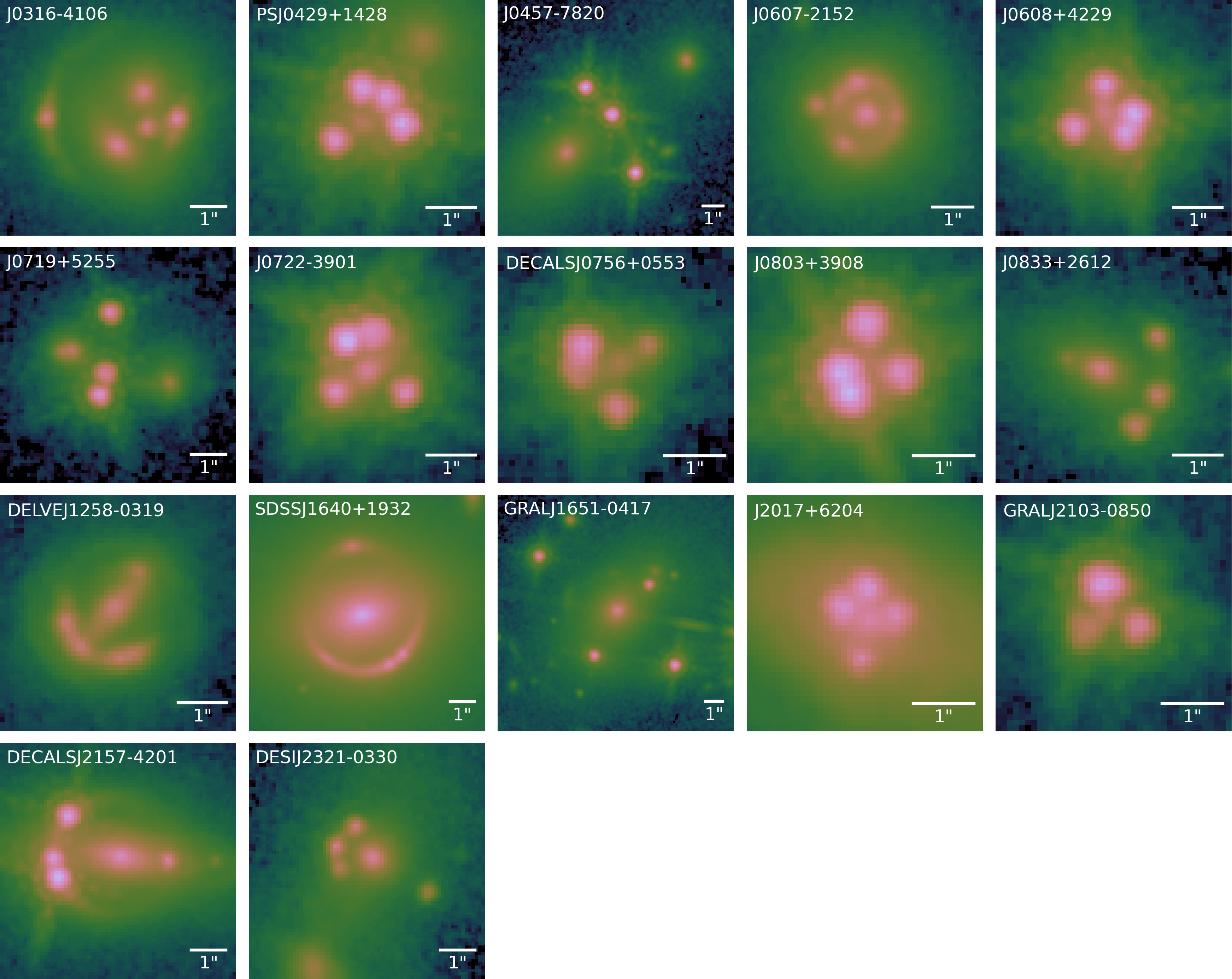}
 \caption{Our 17 lensed quasar sample, displayed in the \textit{HST} F160W IR band.  Our sample includes two triply-lensed, and 15 quadruply-lensed systems.  All images are orientated such that up is North, and left is East.  }
 \label{fig:college}
\end{center}
\end{figure*}

\begin{table*}
\begingroup
\renewcommand{\arraystretch}{1.25}
\begin{center}
\caption{Our lensed quasar sample, with columns representing (1) the target full name, (2) the shorthand name, (3) the right ascension in degrees, (4) the declination in degrees, (5) the lens redshift, (6) the source redshift, (7) the number of lensed quasar images, (8) the presence of a second galaxy, and (9) the paper or people credited for the lensed quasar identification.  We add an asterisk ($^*$) at the end of redshifts where the spectroscopic measurements are tentative.  For systems without source and/or lens redshifts, we assume fiducial redshifts of \zlmean and \zsmean, which are the mean redshifts of our known samples.  For column (8), ``D'' represents a dual lens, whereas ``P'' represents a perturber.}\label{tab:fullsample}
\begin{tabular}{cc|ccccccc}
\hline
\multicolumn{2}{c|}{\ \ \ \ \ \ \ \ \ \ \ \ \ \ \ Name} & R.A. [$^\circ$] & Decl. [$^\circ$] & $z_{\rm d}$ & $z_{\rm s}$ & Images & 2nd lens & L-Quasar Identification \\
(1) & (2) & (3) & (4) & (5) & (6) & (7) & (8) & (9) \\
\hline
\hyperref[v1]{J0316-4106} & J0316 & 49.0031 & -41.1027 & 0.686 & 2.647 & 3 & D & A. Agnello \& P. Schechter \\
\hyperref[v3]{PSJ0429+1428} & J0429 & 67.3048 & 14.4779 &  & 3.866 & 4 & P & \citet{Desira_2022} \\
\hyperref[v4]{J0457-7820} & J0457 & 74.3483 & -78.3466 &  & 3.145 & 3 & D & \customcite{Lemon_2022}{L22} \\
\hyperref[v5]{J0607-2152} & J0607 & 91.7954 & -21.8716 & 0.555 & 1.302 & 4 & P & \customcite{Stern_2021}{S21}, \customcite{Lemon_2022}{L22} \\
\hyperref[v6]{J0608+4229} & J0608 & 92.1725 & 42.4935 & 0.997$^*$ & 2.345 & 4 & - & \customcite{Stern_2021}{S21}, \customcite{Lemon_2022}{L22} \\
\hyperref[v7]{J0719+5255} & J0719 & 109.9324 & 52.9316 &  &  & 4 & D & \customcite{Dawes_2023}{D23} \\
\hyperref[v8]{J0722-3901} & J0722 & 110.6774 & -39.0331 &  &  & 4 & - & C. Lemon \\
\hyperref[v9]{DECALSJ0756+0553} & J0756 & 119.1313 & 5.8877 &  &  & 4 & - & \customcite{Dawes_2023}{D23}, \citet{He_2023} \\
\hyperref[v10]{J0803+3908} & J0803 & 120.9906 & 39.1398 & 1.121 & 2.975 & 4 & - & \customcite{Lemon_2022}{L22} \\
\hyperref[v11]{J0833+2612} & J0833 & 128.4767 & 26.2029 &  & 3.26 & 4 & - & \customcite{Lemon_2022}{L22} \\
\hyperref[v12]{DELVEJ1258-0319} & J1258 & 194.7347 & -3.3289 & 0.691$^*$ & 2.225 & 4 & - & \citet{Schechter_2025} \\
\hyperref[v13]{SDSSJ1640+1932} & J1640 & 250.1903 & 19.5492 & 0.195 & 0.778 & 4 & - & \citet{Wang_2017} \\
\hyperref[v14]{GRALJ1651-0417} & J1651 & 252.7724 & -4.2903 & 0.591 & 1.451 & 4 & - & \customcite{Stern_2021}{S21} \\
\hyperref[v15]{J2017+6204} & J2017 & 304.4544 & 62.0787 & 0.201 & 1.724 & 4 & - & \citet{Delchambre_2019} \\
\hyperref[v16]{GRALJ2103-0850} & J2103 & 315.8708 & -8.8469 & 0.768 & 2.455 & 4 & - & \customcite{Stern_2021}{S21} \\
\hyperref[v17]{DECALSJ2157-4201} & J2157 & 329.3064 & -42.0304 &  &  & 4 & - & \citet{Agnello_2019} \\
\hyperref[v2]{DESIJ2321-0330} & J2321 & 350.3458 & -3.5082 &  & 3.165 & 4 & P & \citet{Sheu_2024} \\
\hline
\end{tabular}
\end{center}
\endgroup
\end{table*}

Our sample of 17 lensed quasars were initially selected as a set of recently-discovered systems.  Many of the systems were discovered using a variety of methodologies, such as using photometric and astrometric data from \textit{Gaia} DR2 \citep[][hereafter S21]{gaia2016, Stern_2021, Gaia_search},  Pan-STARRS \citep[][hereafter L22]{panstars, Lemon_2022}, and the DESI Legacy Imaging Surveys \citep[][hereafter D23]{desils, Dawes_2023}.  See Figure~\ref{fig:college} for a montage of our lensed quasar sample in the recently-observed \textit{HST} F160W filter, and Table~\ref{tab:fullsample} for a summary of our sample.  

For many systems, we do not have spectroscopic measurements.  As such, if we do not have an associated lens or quasar redshift for a system, we assume a fiducial redshift of \zlmean and/or \zsmean, which are the mean redshifts of our known samples.  We only use the redshifts of a given system for the estimation of the time delays and time-delay distance uncertainty; all other results presented in Section~\ref{sec:results} are independent of redshifts.  

We visually identify six systems which we anticipate to have an additional second galaxy that contributes an appreciable flexion to our lens model.  For these secondary galaxies, we assume that lie on the same redshift plane as their primary lens galaxy.  We further classify these secondary galaxies as a dual lens or a perturber.  If we find that the image positions and arcs seem to indicate a single primary lens (i.e., exhibiting standard, textbook configurations), we label the secondary galaxy as a perturber.  Otherwise, we classify the label galaxy as a dual lens.  Depending on its classification, we probe the logarithmic slope of the secondary galaxy's power-law profile ($\gamma_{\rm pl}$) or fix it to $\gamma_{\rm pl}=2$ (isothermal).  See Section~\ref{sec:methodology} for the complete picture of the lens model used in our pipeline.  We only identify the two triple-imaged systems (J0316-4106 and J0457-7820) and an nearly-triple-imaged system (J0719+5255) as dual lenses, and three other systems (PSJ0429+1428, J0607-2152, and DESIJ2321-0330) as having a perturber.

\subsection{J0316-4106}\label{v1}
We present an analysis of J0316-4106, a rare, triply imaged lensed quasar first identified by A. Agnello and P. Schechter (private communication).  The system features a dual lens, with three quasar images situated roughly perpendicular to the axis connecting the two lensing galaxies—a configuration referred to as a ``straddled saddle.''  The precise alignment of these galaxies allows the typically demagnified central image to be visible and even magnified.  Because the quasar images lie at similar distances from each lens, both lenses contribute comparably to the effect.  Therefore, we model the system as a true double lens rather than a primary galaxy with a perturber.  The quasar redshift of $z_{\rm s} = 2.647$) was measured from followup NTT observations (taken by Y. Apostolovski, reduced by C. Spiniello); the deflector redshift of $z_{\rm d} = 0.686$ was measured from IMACS followup observations (taken by P. Schechter, reduced by N. Morgan).

\subsection{PSJ0429+1428}\label{v3}
PSJ0429+1428 was discovered and spectroscopically confirmed in \citet{Desira_2022}, with a quasar redshift of $z_{\rm s} = 3.866$.  There is a nearby perturber $2''$ Northwest of the lens.  However, this system exhibits a ``cusp'' image orientation around the primary lens, affirming that the primary lens is the main lensing body.  Therefore, we treat the secondary galaxy as a perturbing body ($\gamma_{\rm pl}=2$). 

\subsection{J0457-7820}\label{v4}
This system was first identified by \customcite{Lemon_2022}{L22}, with followup NTT-EFOSC2 spectra measuring the quasar redshift of $z_{\rm s} = 3.145$.  Due to the presence of two primary lenses with seemingly comparable lensing effects, there are only three lensed quasar images.  For our analysis, we assume that both lenses are at the same redshift, as we do not have any definitive spectroscopic measurements for either.  Effectively, the second lens galaxy extends the ellipticity of the first.  This allows for the internal diamond caustic to extend past the elliptical caustic on the source plane to where the quasar lies, a mechanism completely different from J0316-4106's triple images.  This rare phenomenon is called a ``naked-cusp'' scenario \citep[e.g.,][]{Brada_2004, Sheu_2024b}.  J0457-7820 is briefly discussed by \citet{Keeley_2025} in relation to its quasar flux ratios, but due to its unusual orientation, it was excluded by their analysis.  Of our sample, this is the highest known source redshift system.

\subsection{J0607-2152}\label{v5}
J0607-2152 was discovered by \customcite{Stern_2021}{S21} and later independently discovered by \customcite{Lemon_2022}{L22}, both using \textit{Gaia} DR2.  Both papers independently measure the quasar redshifts at $z_{\rm s} = 1.302$ using the Keck LRIS and William Herschel Telescope ISIS instruments, respectively.  \citet{Keeley_2025} measures the lens redshift at $z_{\rm d} = 0.555$ on a later run with the Keck LRIS, where they also analyze the lensed quasar flux ratios.  There is a small satellite $1\farcs2$ East of the primary lens.  However, as there is the prominent Einstein ring around the primary lens, it is clear that the primary lens is the overwhelmingly dominant lensing effect in the system.  As such, we treat the satellite as a perturber and fix its slope to isothermal.  

\subsection{J0608+4229}\label{v6}
Like J0607-2152, J0608+4229 was co-discovered and spectroscopically confirmed by \customcite{Stern_2021}{S21} and \customcite{Lemon_2022}{L22}, at $z_{\rm s} = 2.345$.  \customcite{Stern_2021}{S21} dubbed this system as ``Auriga’s Slingshot,'' named after its residing constellation.  With the \textit{HST} imaging resolution, it is possible to identify the lens galaxy despite the smaller separation between the ``fold''-orientated quasar images.  From Subaru FOCAS long-slit spectroscopy (PID: S25B-063; PI: K. C. Wong), we identify a tentative deflector redshift of $z_{\rm d} = 0.997$ using Mg and Ca\,\textsc{ii} H and K absorption features, as well as a slight shift in continuum from the 4000\AA\ break.  \citet{Keeley_2025} performed followup analysis on this system to probe the image flux ratios.

\subsection{J0719+5255}\label{v7}
J0719+5255 is a dual-lens system, with three prominent images in-between the two lenses (like with J0457-7820).  However, unlike the ``naked-cusp'' scenario in J0457-7820, this system exhibits a fourth faint quasar image on the opposing side of the primary lens; this observation was only made possible by the resolution provided by \textit{HST}.  This system was initially identified by \customcite{Dawes_2023}{D23} and independently identified as a quad by C. Lemon, but no additional spectroscopy is available.

\subsection{J0722-3901}\label{v8}
While not previously published, J0722-3901 was discovered by C. Lemon from the \textit{Gaia} DR3 Focused Product Release \citep[FRP;][]{Gaia_search}.  This system also exhibits a ``fold'' image orientation in a single-galaxy lens configuration.  We await spectroscopic measurements for J0722-3901, but the imaging data alone is enough to confirm its quadruply lensed quasar status.

\subsection{DECALSJ0756+0553}\label{v9}
DECALSJ0756+0553 is a single lens quad system, co-discovered by \customcite{Dawes_2023}{D23} and \citet{He_2023}. Similar to J0719+5255 and J0722-3901, this system exhibits a ``fold'' image orientation and lacks spectroscopic confirmation, though the imaging data and lens models presented here are conclusive of its strong-lensing status.  

\subsection{J0803+3908}\label{v10}
J0803+3908 was first discovered by \customcite{Lemon_2022}{L22}, where the authors measure a quasar spectroscopic redshift of $z_{\rm s} = 2.975$.  This system is a single galaxy lens in a ``fold'' orientation.  Despite the small separation between images and their large amplitudes, we are able to discern the lens galaxy light visually and within our lens model. From DESI DR1 \citep{desidr1} aperture and Subaru FOCUS long-slit spectroscopy (PID: S25B-063; PI: K. C. Wong), we measure the lens redshift of $z_{\rm d} = 1.121$ from its prominent absorption features, making it the highest redshift known lens in our sample.  This system was also analyzed by \citet{Keeley_2025} to probe its image flux ratios.

\subsection{J0833+2612}\label{v11}
This ``cusp'' system was identified by \customcite{Lemon_2022}{L22}.  In the discovery paper, they also measure a source redshift of $z_{\rm s} = 3.26$.  While the counterimage is faint and near the lensed galaxy, the \textit{HST} F160W filter observation is successfully able to resolve it.

\subsection{DELVEJ1258-0319}\label{v12}
\citet{Schechter_2025} identified DELVEJ1258-0319 using the DELVE survey and WISE photometry.  Unlike most systems in our sample, this lensed quasar is a low-luminosity, Type-2 AGN.  This is evident in the spectra obtained by \citet{Schechter_2025} (at $z_{\rm s} = 2.225$, exhibiting only narrow-line peaks), and by the lack of visually-obvious PSF images in the \textit{HST} image.  Despite the low luminosity, the source images should still be well modeled by point sources.  From DESI DR1 aperture spectroscopy, we measure a lens redshift of $z_{\rm d} = 0.691$ \citep{desidr1}.  This single-lens system exhibits a ``cusp'' image orientation, with striking host galaxy light.

\subsection{SDSSJ1640+1932}\label{v13}
SDSSJ1640+1932 was discovered, identified, and spectroscopically confirmed by \citet{Wang_2017}, where a lens redshift of $z_{\rm d} = 0.195$ and source redshift of $z_{\rm s} = 0.778$ were measured.  Of the known redshifts in our sample, this system has the lowest lens and source redshifts, allowing for a much higher S/N in the host galaxy lensed arcs.  This single-galaxy lens presents a ``fold'' image orientation.

\subsection{GRALJ1651-0417}\label{v14}
GRALJ1651-0417 is, by far, the largest Einstein-radius system in our sample, which we suspect to be part of a group or small cluster.  In that case, the primary lens is expected to be the brightest cluster galaxy, at the center of a large dark-matter halo.  With a maximum separation of $10''$ between quasar images, the system is classified as an Einstein cross.  In \customcite{Stern_2021}{S21}, GRALJ1651-0417 (nicknamed the ``Dragon Kite'') was first identified as a lensed quasar, where they report deflector and source redshifts of $z_{\rm d} = 0.591$ and $z_{\rm s} = 1.451$.  In addition to the quasar and its host galaxy lensed light, we also observe two additional sets of lensed arcs which we suspect to be from source galaxies at varying redshifts.  While a multiplane lensing analysis could prove to be very fruitful for this system \citep[e.g.,][]{Sheu_2024b, odonnell2026, urcelay2026}, we chose to mask out these features for our pipeline analysis presented in this paper.

\subsection{J2017+6204}\label{v15}
First discovered by \citet{Delchambre_2019} and later spectroscopically confirmed by \customcite{Stern_2021}{S21} ($z_{\rm s} = 1.724$).  The lens galaxy exhibits a complex light profile, and from \textit{JWST} MIRI imaging, \cite{Keeley_2025} infers that it is a spiral galaxy.  Nevertheless, we attempt to model this system given our assumptions of the lensing and light profiles, though our lens light profile may not adequately capture the complexity of the galaxy.  \citet{Keeley_2025} measure a lens redshift of $z_{\rm d} = 0.201$, and the quasar images are arranged in an Einstein cross around the lens.

\subsection{GRALJ2103-0850}\label{v16}
This system, nicknamed the ``Aquarius’ Tear,'' was discovered by \customcite{Stern_2021}{S21}, where they measure the lens and quasar redshifts to be $z_{\rm d} = 0.768$ and $z_{\rm s} = 2.455$.  Due to its small separation and the ``fold'' configuration, we see two of its brightest images in very close proximity, which the \textit{HST} IR resolution is just able to resolve as two separate PSFs.  Combined with the fact that the lens light and host galaxy light are faint, we anticipate it to be difficult to obtain tight constraints on the predicted time delay with this data.  

\subsection{DECALSJ2157-4201}\label{v17}
DECALSJ2157-4201 was first discovered by \citet{Agnello_2019}, and later independently identified in \customcite{Dawes_2023}{D23}, \citet{He_2023}, and by C. Lemon, through morphological and color searches within surveys.  While no redshifts have been measured for this system, the imaging data presented here confirms its status as a lensed quasar, given the presence and ``cusp''-like orientation of its images.  

\subsection{DESIJ2321-0330}\label{v2}
DESIJ2321-0330 was first identified as a lensed quasar candidate by \citet{Sheu_2024}, through measuring photometric variability of strong-lensing candidates within the DESI Legacy Imaging Surveys DR9 \citep{desidr9}.  Prior to this, this system was identified by \citet{Storfer2024} as a strong lens candidate.  In the DESI DR1, they find that the lensed quasar lies at $z_{\rm s} = 3.165$ \citep{desidr1}.  There is a perturbing galaxy approximately $3\farcs7$ Northeast of the primary galaxy, effectively extending the ellipticity of the lens (to the first order).  In our modeling scheme, we nevertheless explicitly account for a separate perturbing body.  DESIJ2321-0330 exhibits a classic lensing ``cusp'' scenario despite the perturber.

\begin{figure*}
\begin{center}
 \includegraphics[width=1\linewidth]{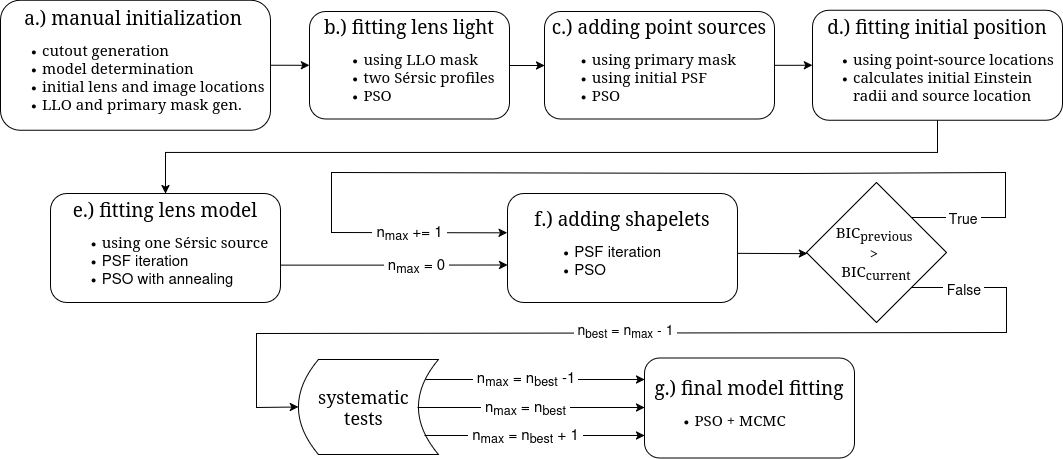}
 \caption{Our modeling block diagram, illustrating the pipeline used for our sample.  Each step is discussed further in detail in Section~\ref{sec:methodology}.  Our modeling pipeline is proven to be robust even against the most unconventional lens systems in our sample.}
 \label{fig:flow_diagram}
\end{center}
\end{figure*}

\section{Methodology} \label{sec:methodology}
To model our sample of 17 lensed quasars, we use the software \textsc{Lenstronomy}\footnote{\url{https://lenstronomy.readthedocs.io/en/latest}} \citep{birrer2018, lenstronomyII} to construct our lens models. \textsc{Lenstronomy} has been shown to be robust in the Time-Delay Lens Modeling Challenge \citep[TDLMC;][]{Ding2021}, where two independent teams used \textsc{Lenstronomy} to recover lens model parameters with statistical consistency in Rung 2.  \textsc{Lenstronomy} has since been instrumental in providing reliable lens modeling capabilities within the TDCOSMO collaboration, and has been extensively tested both internally and externally \citep[e.g.,][]{Shajib_2022, Schmidt_2022, Ertl_2023}.  In Figure~\ref{fig:flow_diagram}, we illustrate our full modeling pipeline as a block diagram, whihc we go into detail in the following subsection.

\subsection{PSF generation}
As we are modeling lensed quasars, understanding the PSF is paramount.  In our analysis, we iteratively fit the PSF at multiple steps throughout our modeling pipeline.  Our initial PSF profile is generated from stacking nearby stars across all of our observations.  While ideally we would generate an individual initial PSF for each observation based on stars within their own respective fields, some systems reside in much sparser fields than others, resulting in poor PSF fits.  By normalizing and stacking 106 stars across all 17 fields, we generate a detailed starting PSF, with the variance map calculated from the pixel-wise scatter across the sample of stars \citep{lenstronomyII}.

\subsection{Modeling pipeline}
Our pipeline begins with a manual initialization for each system.  This involves generating the image cutout (i.e., its size and positioning), determining the starting positions of the lens(es) and image locations, and creating a mask to hide extraneous light from the lens model.  In addition to a primary mask used for most of the pipeline, we also generate a lens-light-only (LLO) mask to generously hide the source and quasar light, so as to show only the lens galaxy light.  This LLO mask is utilized in the next step when we fit for the lens light profiles only.  Also in the initialization, we decide which systems have a secondary galaxy, and whether it is considered a dual lens or a perturbing galaxy (as discussed in Section~\ref{sec:quasarsample}).

Next, with \textsc{Lenstronomy} and using the starting locations of the lens position(s), we fit for only the lens galaxies' light profiles using the LLO mask.  For each lens galaxy, we use two Sérsic profiles to represent the galaxy bulge and halo light profiles.  The two Sérsic profiles are jointly constrained such that they share the same centroid and ellipticity angle $\phi$.  They are also modeled such that the Sérsic component representing the bulge has a steeper profile and a smaller half-light radius compared to the Sérsic profile corresponding to the halo (i.e., $n_{\rm b} > n_{\rm h}$ and $R_{\rm b} < R_{\rm h}$).  Rather than simply using de Vaucouleurs \citep[$n_{\rm b} = 4$,][]{vaucouleurs1948} and exponential ($n_{\rm h} = 1$) light profiles, we find that our systems necessitate a more complex light profile in the IR F160W filter.  We use a particle swarm optimization (PSO) algorithm to fit the lens light parameters here.

We then add the quasar point sources to our model.  Using the primary mask from this point forward, we utilize the initial positions of the quasar images and the initial PSF to fit the combined light profiles of the lens galaxies and quasar images.  At this point, we are still not introducing any lensing to our model; we are only photometrically fitting the images.  We also do not iterate over our PSF, as we expect that there is still considerable host galaxy light we have not taken into account.  The main purpose of this step is to get a reasonable first estimate of the positions of the quasar images.  We run PSO to fit our current model at this step.

Once we have a good estimate of the quasar image positions and the shape of the lens light, we perform a position-based fit to place initial constraints on the lensing parameters.  Here, fiducially assuming the ellipticity of the bulge light and assuming $\gamma_{\rm pl} = 2$, we estimate the Einstein radius that minimizes the scatter of the image positions when ray-traced back to the source plane.  If a secondary galaxy is present, we fit for both Einstein radii and follow the same procedure.  With this method, we can obtain a preliminary yet robust measurement of the Einstein radii and the initial position of the source quasar (derived from the mean of the ray-traced image positions).  

Then we finally add lensing effects to our model, using the previous step's results as a starting point.  For each applicable lens galaxy, we add an elliptical power law to the model, with an additional external shear component.  For each power law profile, we add priors such that it agrees well with the light profile; these priors include a Gaussian prior on the centroid position of $\sigma=0\farcs1$, a Gaussian prior on the ellipticity angle $\phi$ discrepancy between the light components and the lens profile of $\sigma=10^{\circ}$, and an axis ratio $q$ bound of $q_{\rm mass} < q_{\rm light}$ on both the bulge and halo components.  For the source light profile, we use an exponential light profile, with the quasar point source fixed to the center of the profile.  We opt to use an exponential profile over the more flexible Sérsic since we introduce additional source complexity in the form of shapelets in the next step.  We include Gaussian priors on the source plane to ensure that the lensed image positions, when traced back to the source plane, are close ($\sigma=0\farcs0003$) to the center of the exponential light profile.  We also allow a deviation from the predicted quasar image position on the image plane (with a prior of $\sigma=0\farcs001$) for each image to account for slight instrumental or systematic differences.  We use PSO to fit the lens model, while iteratively fitting the PSF throughout the process. 

Starting at a maximum shapelets basis order of $n_{\rm max}=0$, we fit our model and measure the Bayesian information criterion (BIC) of the best-fitting parameterization.  If we find that the BIC is lower than that of the previous $n_{\rm max}$ model, we rerun the model with $n_{\rm max, current} = n_{\rm max, previous} + 1$.  From this methodology, we use the BIC as a heuristic to deduce what shapelets order basis works best for the data.  For each of the different $n_{\rm max}$ runs, we iteratively fit the PSF in conjunction with PSO for the full model.  

When we achieve a $n_{\rm max, best}$ setting such that the BIC increases with a higher value, we then split the model up into three systematic tests: where $n_{\rm max} = n_{\rm max, best}$, $n_{\rm max} = n_{\rm max, best} - 1$, and $n_{\rm max} = n_{\rm max, best} + 1$.  For each of these parameterizations, we run a PSO fit followed by a Markov chain Monte Carlo (MCMC) algorithm to probe the final posteriors.  The chains are run until all parameters have converged to a stable distribution.  The final results are then generated from the BIC-weighted marginalization across all three systematic models.

Our pipeline draws heavy inspiration from previously done \textsc{Lenstronomy}-based lensed-quasar pipelines, such as \citet{shajib2018} and \citet{Schmidt_2022}.  However, our circumstances are notably different compared to the two other papers.  Primarily, we only have \textit{HST} F160W observations from our \textit{HST} program, whereas \citet{shajib2018} and \citet{Schmidt_2022} built their pipelines to model \textit{HST} F475X, F814W, and F160W filters.  While the optical F475X and F814W offer superior resolution and astrometry measurements, the F160W filter often stronger source light emission, which contributes to tighter constraints on the lens model.  The F160W observations, however, display significantly more complex lens-light profiles.  In \citet{shajib2018} and \citet{Schmidt_2022}, this resulted in worse fits in the the F160W compared to the optical bands, but the overall fit is nevertheless excellent.  As we do not have the optical bands, we place heavier emphasis on the lens-light profile modeling, allowing for a wider parameter space.  This is achieved by not fixing the Sérsic indices of the bulge and halo components, providing informative priors to ensure they remain physical, and additional components to the light profile when necessary (see Section~\ref{subsec:sys_exemptions}).  

\subsection{System exceptions} \label{subsec:sys_exemptions}
For system GRALJ1651-0417, an extremely bright star is within the cutout field, contributing significant excess flux.  Unfortunately, one of the star's PSF spikes goes directly through the lens galaxy.  To account for these issues, we model and subtract out the star flux from our model, before running the full aforementioned pipeline.  We also mask a majority of the PSF spike, except near the lens galaxy where we expect the signal from the lens far exceeds that of the spike.  Following these preprocessing steps, we are able to recover a well-fit model.

After running our pipeline on the full sample, we found that for systems J0607-2152, J0833+2612, SDSSJ1640+1932, and DESIJ2321-0330, the light profile residuals were significant enough to warrant an additional component.  For these systems, we find that adding a point source component at the center of the primary lens galaxy results, resulted in an adequate fit and reduced residuals.  This is necessary given the complexity of the lens galaxy and the IR imaging granted by the F160W filter, likely indicating a dense lens galaxy core.  

\section{Results} \label{sec:results}

In Tables~\ref{tab:summary_primary} and \ref{tab:summary_secondary}, we provide the summary parameters describing the primary and secondary (if applicable) galaxies, respectively.  We put additional modeling results in Appendix~\ref{app:A}. There, we show the best-fit model plots in Figures~\ref{fig:model_figure0}, \ref{fig:model_figure1}, and \ref{fig:model_figure2}, as well as the full parameterizations of the light profiles in Tables~\ref{tab:primary_bulge}, \ref{tab:primary_halo}, \ref{tab:secondary_bulge}, and \ref{tab:secondary_halo}.  Additionally, in Figure~\ref{fig:college_source}, we present the lens-light and point-source subtracted data image, to illustrate the amount of lensing information (i.e., extended emission from the source) within each system \citep{Tan_2024, Sheu_2025_dinos}.  Finally, we provide the convergence and shear values for each quasar image of each system in Table~\ref{tab:kappa_gamma}.  This is particularly useful for future microlensing studies.  

We find that most of the systems resulted in satisfactory models from the pipeline.  However, we find that our pipeline struggles on a few lenses, where the lens galaxy light is more complex than what our model allows.  DELVEJ1258-0319, J2017+6204, and GRALJ2103-0850 have correlated residuals which we attribute to the spiral-galaxy-like features in their light profiles.  Therefore, we posit that for those three systems, their primary lenses are spiral galaxies.  \citet{Keeley_2025} also come to a similar conclusion from their visual inspection of their lens sample (of which we overlap).  For J0316-4106 and DESIJ2321-0330, we find that there is detectable mixing between their primary and secondary galaxies, resulting in fairly minor distortions in their light profiles.  Lastly, SDSSJ1640+1932 displays a highly elliptical light profile with, from the residuals in Figure~\ref{fig:model_figure1}, a clear necessity for additional multipole moments.  We expect that a multipole order of $m=4$ is present, given the symmetry in the residual image.  Since there is currently no implementation of a multipole-perturbed S\`ersic profile in \textsc{Lenstronomy}, we attempt to achieve a better fit with an additional PSF at the core.  While all of these systems discussed have minor to major lens light modeling difficulties, we nevertheless provide these models as a good first step for future studies to build on. 

In Table~\ref{tab:tds}, we provide our model-measured Fermat potential differences and time delays between images, using image A as the reference.  See Figures~\ref{fig:model_figure0}, \ref{fig:model_figure1}, and \ref{fig:model_figure2} for our image naming scheme for each system.  In our time delay determination, we assume a fiducial cosmology of $H_0=70$ km~s$^{-1}$~Mpc$^{-1}$ and $\Omega_{\rm m, 0}=0.3$.  The lens and source redshifts are also necessary for the calculation, which we take from Table~\ref{tab:fullsample}.  For systems without a $z_{\rm d}$ and/or $z_{\rm s}$ measurement, we assume the mean lens and/or source redshifts of our sample: \zlmean and \zsmean.  In Table~\ref{tab:mags}, we provide the lensing magnifications and the apparent magnitudes of each image from our pipeline modeling.  As the time-delay uncertainty is directly correlated to how well we can constrain $H_0$, our time-delay predictions and photometry measurements will inform future cosmography analysis on these systems.

For the triply-imaged systems J0316-4106 and J0457-7820 (as well as for the near-triply-imaged J0719+5255), our pipeline is able to capture their atypical lensing configuration (see Figure~\ref{fig:model_figure0}).  

Finally, we note that there are measured time-delay measurements for GRALJ1651-0417 by \citet{gral16_tds}, using the Zwicky Transient Facility.  They identify time delays that are approximately three times larger than what we predict, as we find that our model prefers a significantly lower $\gamma_{\rm pl}\sim1.2$ which results in a significantly lower time delay predictions. As a reminder, lens modeling primarily constrains the slope of the mass density profile around the Einstein radius (i.e., at the location of the lensed images of the extended source).  As we are assuming a single-power-law mass profile, this constraint is then imposed across the full radial range. This is usually a safe assumption in typical massive elliptical galaxies \citep[i.e., the ``bulge-halo conspiracy''; see][]{Dutton_2014, Etherington2023}.  However, as GRALJ1651-0417 is nearing a massive group/cluster scale, where this assumption fails \citep{Newman2015}.  As such, we recommend that future work regarding GRALJ1651-0417 utilizes a more complex lens model (e.g., an NFW + Sérsic lensing profile), which is out-of-scope for this pipeline-modeling paper analysis.

\subsection{Estimating the time-delay distance uncertainty} \label{subsec:obs_td_unc}
To assess whether a system has a strong potential to constrain $H_0$, we attempt to quantify the relative uncertainty that can be obtained on the time delay distance based on the models and potential time delay precision from hypothetical state of the art monitoring campaigs. \citet{Schmidt_2022} does this by probing the stability of the Fermat potential differences between different models.  In this paper, we will take this a step further by incorporating an observational uncertainty component, and aggregating the combined time-delay distance constraining power of a given system.  From Equation~\ref{eq:td}, the uncertainties on the time delays ($\sigma_{\Delta t}$) and the Fermat potential ($\sigma_{\Delta \Phi}$) are added in quadrature for the uncertainty on $D_{\Delta t}$; the former being the observational component and the latter being the modeling uncertainty component.  

With current time-delay measurement techniques, the best one can achieve is an error of one to two days, driven by many factors such as (but not limited to) the cadence of observations, the variability of the quasar, microlensing effects, and the time-frame of observations \citep{Dux_2025}.  Fiducially, we assume a $\sigma_{\Delta t} = 2 {\rm ~days}$ to represent the time-delay measurement uncertainty for a given pair of images attainable with a state of the art ground based monitoring campaign.  The modeling-uncertainty component is provided by the uncertainties of $\Delta\Phi$ and $\Delta t$ in Table~\ref{tab:tds}.  As the two values are simply scaled by a $D_{\Delta t}$ (defined by our fiducial cosmology) divided by the speed of light (Equation~\ref{eq:td}), they share the same relative uncertainties.  However, to be aggregated with the observational uncertainty in days, $\sigma_{\Delta \Phi}$ must also be in the corresponding units of days, and so we must assume a cosmology for this measurement ($H_0=70$ km~s$^{-1}$~Mpc$^{-1}$ and $\Omega_{\rm m, 0}=0.3$).

Assuming independent Gaussian errors, the combined time-delay distance uncertainty of a system $\sigma_{D \Delta t, \%}$ is defined as:
\begin{equation}
    \frac{1}{\sigma_{D \Delta t, \%}} = \sqrt{\sum_{i \neq b}^{N}  \frac{\mu_{\Delta t, b, i}^2}{\sigma_{\Delta t, b, i}^2+\sigma_{\Delta \Phi, b, i}^2} } ,
\end{equation}
where $N$ is the total number of lensed images, $b$ is the baseline image used as the reference for the time delay, and $\mu_{\Delta t, b, i}$ is the mean time delay between image $b$ and $i$, respectively.  Using this formulation, we can see that the total percent uncertainty will always be better than the best percent uncertainty between the baseline and any given lensed image.  $\mu_{\Delta t, b, i}$ is estimated by our model in Table~\ref{tab:tds}.  We use the quasar image that results in the lowest $\sigma_{D \Delta t, \%}$ as the baseline for each system.  In the last column of Table~\ref{tab:tds} are our values of the time-delay percent uncertainties for each system. 


We group our sample of 17 lensed quasars into four categories, based on their $\sigma_{D \Delta t, \%}$: ``excellent'' ($\sigma_{D \Delta t} \leq 3\%$), ``good'' ($3\% < \sigma_{D \Delta t} \leq 7\% $), ``suitable'' ($7\% < \sigma_{D \Delta t} \leq 12\%$), and ``impractical'' ($\sigma_{D \Delta t} > 12 \%$) for time-delay cosmography. The thresholds are chosen to put the time delay and Fermat potential terms of the error budget in the context of to other terms such as residual MST from stellar kinematics and line of sight convergence, which are of order a few percent. For the "excellent" systems the term estimated here will be subdominant compared to other sources of error, for the "good" systems it will comparable, for the "suitable" ones it will be the dominant term, and for the "impractical" ones this term alone is sufficient to make the system not worthy of a dedicated monitoring campaign.

There are six systems in the ``excellent'' category (J0316-4106, J0719+5255, SDSSJ1640+1932, GRALJ1651-0417, DECALSJ2157-4201, and DESIJ2321-0330), indicating the potential for a sub-percent-level measurement of their time delay.  Though some of their redshifts remain unknown, they nevertheless show a strong likelihood for significant time-delay cosmography.  There are five systems in the ``good'' category (J0457-7820, J0608+4229, J0803+3908, J0833+2612, and DELVEJ1258-0319), where many of the current TDCOSMO lenses are as well.  There are three systems in the ``suitable'' category (J0607-2152, J0722-3901, and DECALSJ0756+0553), and finally three in the ``impractical'' category (PSJ0429+1428, J2017+6204, and GRALJ2103-0850).  

We note that our estimation leans on assumptions of the redshifts (if unknown), idealized time-delay measurements, and modeling uncertainties based on one \textit{HST} filter.  Nevertheless, we assert that it is a good heuristic for whether a system is worth investing time-intensive observations for the sake of $H_0$ cosmography.


\begin{table*}
\begingroup
\renewcommand{\arraystretch}{1.25}
\begin{center}
\caption{Summary pipeline results for the primary lens profiles plus external shear.  $\theta_{\rm E, p}$ is the Einstein radius, $\gamma_{\rm pl, p}$ is the logarithmic slope of the power-law mass profile, $q_{\rm m, p}$ is the semi-major to semi-minor axis ratio of the mass profile, $\phi_{\rm m, p}$ is the East-of-North position angle of the mass profile, $R_{\rm eff, p}$ is the circular effective radius of the light profile, $m_{\rm p}$ is the $2''$ aperture apparent magnitude of the lens galaxy modeled light, $\gamma_{\rm ext}$ is the external shear contribution, and $\phi_{\rm ext}$ is the East-of-North position angle of the external shear.  All parameters shown, with exemption to $\gamma_{\rm ext}$ and $\phi_{\rm ext}$, pertain to the primary lens.}\label{tab:summary_primary}
\begin{tabular}{c|cccc:cc:cc}
\hline
Name & $\theta_{\rm E, p}$ & $\gamma_{\rm pl, p}$ & $q_{\rm m, p}$ & $\phi_{\rm m, p}$  & $R_{\rm eff, p}$ & $m_{\rm p}$ & $\gamma_{\rm ext}$ & $\phi_{\rm ext}$ \\
 & [$''$] &   &   & [$^\circ$]  & [$''$] & [mag] & & [$^\circ$] \\
\hline
J0316 & $0.760^{+0.010}_{-0.009}$ & $1.95^{+0.01}_{-0.01}$ & $0.595^{+0.004}_{-0.003}$ & $45.1^{+0.7}_{-0.6}$ & $0.772^{+0.006}_{-0.023}$ & $18.94^{+0.01}_{-0.01}$ & $0.099^{+0.003}_{-0.003}$ & $62.4^{+0.8}_{-0.8}$  \\ 
J0429 & $0.679^{+0.003}_{-0.003}$ & $2.01^{+0.02}_{-0.03}$ & $0.638^{+0.011}_{-0.012}$ & $-14.0^{+1.1}_{-0.9}$ & $0.617^{+0.047}_{-0.042}$ & $20.51^{+0.02}_{-0.02}$ & $0.114^{+0.004}_{-0.006}$ & $-18.5^{+1.9}_{-1.6}$  \\ 
J0457 & $2.177^{+0.012}_{-0.009}$ & $2.31^{+0.01}_{-0.01}$ & $0.835^{+0.007}_{-0.007}$ & $-28.7^{+0.7}_{-0.6}$ & $2.075^{+0.043}_{-0.043}$ & $19.03^{+0.01}_{-0.01}$ & $0.238^{+0.001}_{-0.002}$ & $61.6^{+0.1}_{-0.1}$  \\ 
J0607 & $0.759^{+0.003}_{-0.003}$ & $2.37^{+0.05}_{-0.04}$ & $0.842^{+0.012}_{-0.011}$ & $53.3^{+1.6}_{-1.5}$ & $0.530^{+0.005}_{-0.005}$ & $18.76^{+0.03}_{-0.04}$ & $0.028^{+0.004}_{-0.004}$ & $-72.5^{+3.4}_{-3.6}$  \\ 
J0608 & $0.635^{+0.002}_{-0.002}$ & $2.44^{+0.05}_{-0.06}$ & $0.527^{+0.033}_{-0.027}$ & $64.7^{+0.9}_{-0.8}$ & $0.285^{+0.017}_{-0.001}$ & $19.55^{+0.01}_{-0.01}$ & $0.025^{+0.005}_{-0.005}$ & $33.1^{+8.9}_{-9.5}$  \\ 
J0719 & $0.783^{+0.010}_{-0.011}$ & $1.74^{+0.02}_{-0.02}$ & $0.994^{+0.004}_{-0.008}$ & $10.7^{+3.8}_{-4.8}$ & $0.326^{+0.022}_{-0.024}$ & $20.41^{+0.04}_{-0.04}$ & $0.071^{+0.004}_{-0.003}$ & $73.2^{+2.3}_{-1.9}$  \\ 
J0722 & $0.819^{+0.002}_{-0.001}$ & $1.99^{+0.05}_{-0.05}$ & $0.687^{+0.013}_{-0.013}$ & $-43.5^{+0.5}_{-0.5}$ & $0.326^{+0.001}_{-0.008}$ & $19.03^{+0.01}_{-0.01}$ & $0.062^{+0.005}_{-0.005}$ & $-32.5^{+1.7}_{-1.4}$  \\ 
J0756 & $0.654^{+0.002}_{-0.002}$ & $2.21^{+0.09}_{-0.05}$ & $0.834^{+0.019}_{-0.023}$ & $-30.1^{+1.8}_{-1.9}$ & $0.285^{+0.001}_{-0.009}$ & $20.77^{+0.02}_{-0.02}$ & $0.060^{+0.009}_{-0.008}$ & $72.2^{+2.8}_{-1.9}$  \\ 
J0803 & $0.575^{+0.001}_{-0.001}$ & $2.39^{+0.03}_{-0.02}$ & $0.986^{+0.008}_{-0.011}$ & $-60.9^{+4.6}_{-4.6}$ & $0.226^{+0.011}_{-0.036}$ & $21.10^{+0.05}_{-0.07}$ & $0.208^{+0.002}_{-0.004}$ & $-80.9^{+0.3}_{-0.5}$  \\ 
J0833 & $1.149^{+0.005}_{-0.004}$ & $2.00^{+0.05}_{-0.05}$ & $0.640^{+0.014}_{-0.014}$ & $31.0^{+0.6}_{-0.7}$ & $0.526^{+0.014}_{-0.029}$ & $19.55^{+0.03}_{-0.02}$ & $0.085^{+0.008}_{-0.008}$ & $80.4^{+3.1}_{-3.9}$  \\ 
J1258 & $0.982^{+0.002}_{-0.002}$ & $1.85^{+0.02}_{-0.02}$ & $0.672^{+0.011}_{-0.007}$ & $-229.4^{+0.4}_{-0.4}$ & $0.326^{+0.001}_{-0.001}$ & $19.45^{+0.01}_{-0.01}$ & $0.058^{+0.005}_{-0.007}$ & $-33.9^{+3.0}_{-2.0}$  \\ 
J1640 & $2.517^{+0.001}_{-0.001}$ & $2.65^{+0.01}_{-0.01}$ & $0.602^{+0.004}_{-0.004}$ & $-197.5^{+0.2}_{-0.2}$ & $2.785^{+0.016}_{-0.039}$ & $16.14^{+0.01}_{-0.01}$ & $0.044^{+0.001}_{-0.001}$ & $89.0^{+0.4}_{-0.4}$  \\ 
J1651 & $3.741^{+0.004}_{-0.004}$ & $1.21^{+0.01}_{-0.01}$ & $0.848^{+0.003}_{-0.003}$ & $-224.5^{+0.3}_{-0.3}$ & $2.192^{+0.080}_{-0.064}$ & $18.44^{+0.01}_{-0.01}$ & $0.018^{+0.001}_{-0.001}$ & $-7.2^{+5.1}_{-5.2}$  \\ 
J2017 & $0.507^{+0.011}_{-0.004}$ & $2.74^{+0.19}_{-0.20}$ & $0.983^{+0.011}_{-0.015}$ & $-160.9^{+4.5}_{-4.6}$ & $1.333^{+0.015}_{-0.008}$ & $18.23^{+0.02}_{-0.01}$ & $0.175^{+0.016}_{-0.019}$ & $-82.4^{+0.5}_{-0.6}$  \\ 
J2103 & $0.504^{+0.002}_{-0.002}$ & $2.34^{+0.05}_{-0.05}$ & $0.671^{+0.028}_{-0.029}$ & $-37.4^{+1.7}_{-1.5}$ & $0.137^{+0.009}_{-0.010}$ & $20.76^{+0.04}_{-0.04}$ & $0.068^{+0.010}_{-0.010}$ & $24.2^{+4.3}_{-4.9}$  \\ 
J2157 & $1.688^{+0.001}_{-0.001}$ & $2.39^{+0.01}_{-0.02}$ & $0.499^{+0.004}_{-0.004}$ & $7.2^{+0.2}_{-0.2}$ & $0.933^{+0.001}_{-0.005}$ & $17.82^{+0.01}_{-0.01}$ & $0.042^{+0.003}_{-0.002}$ & $19.7^{+1.1}_{-1.1}$  \\ 
J2321 & $1.113^{+0.008}_{-0.011}$ & $2.34^{+0.05}_{-0.05}$ & $1.000^{+0.001}_{-0.001}$ & $113.4^{+5.2}_{-5.6}$ & $2.730^{+0.189}_{-0.141}$ & $19.12^{+0.01}_{-0.01}$ & $0.098^{+0.007}_{-0.010}$ & $27.6^{+0.7}_{-0.7}$  \\ 
\hline
\end{tabular}
\end{center}
\endgroup
\end{table*}

\begin{table*}
\begingroup
\renewcommand{\arraystretch}{1.25}
\begin{center}
\caption{Summary pipeline results for the secondary lens profiles.  $\theta_{\rm E, s}$ is the Einstein radius, $\gamma_{\rm pl, s}$ is the logarithmic slope of the power-law mass profile, $q_{\rm m, s}$ is the semi-major to semi-minor axis ratio of the mass profile, $\phi_{\rm m, s}$ is the East-of-North position angle of the mass profile, $R_{\rm eff, s}$ is the circular effective radius of the light profile, and $m_{\rm s}$ is the $2''$ aperture apparent magnitude of the lens galaxy modeled light.  All parameters shown pertain to the secondary lens.}\label{tab:summary_secondary}
\begin{tabular}{c|cccc:cc}
\hline
Name & $\theta_{\rm E, s}$ & $\gamma_{\rm pl, s}$ & $q_{\rm m, s}$ & $\phi_{\rm m, s}$ & $R_{\rm eff, s}$ & $m_{\rm s}$ \\
 & [$''$] &   &   & [$^\circ$]  & [$''$] & [mag] \\
\hline
J0316 & $1.065^{+0.014}_{-0.015}$ & $1.52^{+0.01}_{-0.01}$ & $0.733^{+0.007}_{-0.007}$ & $49.58^{+0.48}_{-0.60}$ & $0.828^{+0.021}_{-0.012}$ & $19.22^{+0.01}_{-0.01}$  \\ 
J0429 & $0.264^{+0.017}_{-0.018}$ & $2$ & $0.993^{+0.005}_{-0.007}$ & $85.88^{+15.80}_{-26.75}$ & $0.785^{+0.013}_{-0.010}$ & $20.21^{+0.01}_{-0.01}$  \\ 
J0457 & $1.097^{+0.035}_{-0.031}$ & $1.99^{+0.03}_{-0.02}$ & $0.867^{+0.016}_{-0.009}$ & $-82.04^{+3.12}_{-3.21}$ & $7.410^{+0.327}_{-0.422}$ & $20.11^{+0.01}_{-0.01}$  \\ 
J0607 & $0.185^{+0.009}_{-0.009}$ & $2$ & $0.977^{+0.017}_{-0.030}$ & $33.82^{+4.67}_{-4.71}$ & $3.582^{+0.638}_{-0.748}$ & $20.58^{+0.04}_{-0.05}$  \\ 
J0719 & $0.928^{+0.019}_{-0.017}$ & $2.00^{+0.02}_{-0.02}$ & $0.796^{+0.031}_{-0.030}$ & $-231.74^{+1.89}_{-1.62}$ & $0.476^{+0.024}_{-0.033}$ & $20.76^{+0.03}_{-0.03}$  \\ 
J2321 & $0.785^{+0.067}_{-0.037}$ & $2$ & $0.979^{+0.016}_{-0.023}$ & $89.77^{+3.71}_{-3.58}$ & $0.660^{+0.015}_{-0.015}$ & $20.01^{+0.01}_{-0.01}$  \\ 
\hline
\end{tabular}
\end{center}
\endgroup
\end{table*}


\begin{table*}
\begingroup
\renewcommand{\arraystretch}{1.25}
\begin{center}
\caption{Modeled Fermat potential differences ($\Delta\Phi$) and predicted time delays ($\Delta t$), with respect to image A.  The estimated time-delay distance uncertainty ($\sigma_{D \Delta t}$) is shown in the last column; see Section~\ref{subsec:obs_td_unc} for details on how this is calculated.  The time delays are calculated assuming a fiducial cosmology of $H_0=70$ km~s$^{-1}$~Mpc$^{-1}$ and $\Omega_{\rm m, 0}=0.3$, and using the redshifts shown in Table~\ref{tab:fullsample}.  If a redshift has not yet been measured, we assume \zlmean and/or \zsmean, which are the mean redshifts of the of our known sample.  These systems have been labeled with an asterisk ($^{*}$) and/or a dagger ($^{\dagger}$) superscript, respectively.  For the identifications of each image for each system, see Figures~\ref{fig:model_figure0}, \ref{fig:model_figure1}, and \ref{fig:model_figure2}.  }\label{tab:tds}
\begin{tabular}{c|ccc:ccc:c}
\hline
Name & $\Delta\Phi_{\rm AB}$ & $\Delta\Phi_{\rm AC}$ & $\Delta\Phi_{\rm AD}$ & $\Delta t_{\rm AB}$ & $\Delta t_{\rm AC}$ & $\Delta t_{\rm AD}$ & $\sigma_{D \Delta t}$ \\
 & & & & [days] & [days] & [days] & [\%] \\
\hline
J0316 & $0.015^{+0.001}_{-0.001}$ & $-1.146^{+0.020}_{-0.018}$ &  & $1.75^{+0.08}_{-0.10}$ & $-134.38^{+2.34}_{-2.10}$ &  & 1.57  \\ 
J0429$^{*}$ & $0.009^{+0.001}_{-0.001}$ & $0.004^{+0.001}_{-0.001}$ & $0.093^{+0.005}_{-0.005}$ & $0.85^{+0.06}_{-0.06}$ & $0.38^{+0.04}_{-0.03}$ & $9.11^{+0.48}_{-0.49}$ & 13.63  \\ 
J0457$^{*}$ & $0.526^{+0.007}_{-0.007}$ & $0.393^{+0.005}_{-0.005}$ &  & $54.05^{+0.70}_{-0.70}$ & $40.38^{+0.53}_{-0.52}$ &  & 3.11  \\ 
J0607 & $-0.058^{+0.003}_{-0.003}$ & $-0.052^{+0.003}_{-0.003}$ & $-0.161^{+0.005}_{-0.006}$ & $-7.02^{+0.39}_{-0.40}$ & $-6.20^{+0.36}_{-0.38}$ & $-19.38^{+0.61}_{-0.67}$ & 7.85  \\ 
J0608 & $0.004^{+0.001}_{-0.001}$ & $-0.095^{+0.006}_{-0.007}$ & $0.026^{+0.002}_{-0.001}$ & $0.82^{+0.10}_{-0.11}$ & $-21.11^{+1.44}_{-1.50}$ & $5.87^{+0.36}_{-0.31}$ & 6.22  \\ 
J0719$^{* \dagger}$ & $0.198^{+0.006}_{-0.004}$ & $0.184^{+0.005}_{-0.004}$ & $0.653^{+0.016}_{-0.015}$ & $22.01^{+0.61}_{-0.42}$ & $20.50^{+0.58}_{-0.39}$ & $72.60^{+1.79}_{-1.61}$ & 2.45  \\ 
J0722$^{* \dagger}$ & $0.087^{+0.005}_{-0.005}$ & $0.085^{+0.005}_{-0.005}$ & $0.113^{+0.006}_{-0.006}$ & $9.72^{+0.55}_{-0.57}$ & $9.39^{+0.55}_{-0.54}$ & $12.58^{+0.70}_{-0.70}$ & 11.35  \\ 
J0756$^{* \dagger}$ & $0.002^{+0.001}_{-0.001}$ & $-0.025^{+0.002}_{-0.001}$ & $0.095^{+0.005}_{-0.006}$ & $0.20^{+0.04}_{-0.04}$ & $-2.72^{+0.17}_{-0.16}$ & $10.55^{+0.57}_{-0.68}$ & 10.62  \\ 
J0803 & $0.049^{+0.001}_{-0.001}$ & $0.045^{+0.001}_{-0.001}$ & $0.174^{+0.003}_{-0.004}$ & $11.27^{+0.24}_{-0.25}$ & $10.52^{+0.21}_{-0.24}$ & $40.32^{+0.68}_{-0.86}$ & 3.69  \\ 
J0833$^{*}$ & $0.040^{+0.002}_{-0.002}$ & $0.024^{+0.001}_{-0.001}$ & $0.671^{+0.032}_{-0.034}$ & $4.08^{+0.22}_{-0.25}$ & $2.44^{+0.14}_{-0.15}$ & $68.33^{+3.27}_{-3.44}$ & 3.34  \\ 
J1258 & $0.006^{+0.001}_{-0.001}$ & $-0.008^{+0.001}_{-0.001}$ & $0.270^{+0.008}_{-0.008}$ & $0.74^{+0.06}_{-0.07}$ & $-0.96^{+0.08}_{-0.07}$ & $34.15^{+1.01}_{-0.99}$ & 3.76  \\ 
J1640 & $-0.003^{+0.002}_{-0.002}$ & $0.104^{+0.001}_{-0.001}$ & $-2.067^{+0.011}_{-0.011}$ & $-0.09^{+0.07}_{-0.07}$ & $3.29^{+0.04}_{-0.04}$ & $-65.21^{+0.33}_{-0.34}$ & 1.77  \\ 
J1651 & $-1.739^{+0.062}_{-0.058}$ & $-0.256^{+0.009}_{-0.008}$ & $-0.804^{+0.030}_{-0.028}$ & $-216.44^{+7.70}_{-7.27}$ & $-31.82^{+1.11}_{-1.03}$ & $-100.10^{+3.67}_{-3.44}$ & 2.13  \\ 
J2017 & $-0.036^{+0.004}_{-0.006}$ & $-0.022^{+0.003}_{-0.006}$ & $-0.087^{+0.008}_{-0.008}$ & $-1.01^{+0.12}_{-0.17}$ & $-0.61^{+0.09}_{-0.16}$ & $-2.42^{+0.21}_{-0.22}$ & 60.27  \\ 
J2103 & $0.037^{+0.002}_{-0.003}$ & $0.036^{+0.002}_{-0.003}$ & $0.091^{+0.007}_{-0.008}$ & $5.18^{+0.34}_{-0.40}$ & $5.06^{+0.34}_{-0.40}$ & $12.95^{+1.01}_{-1.17}$ & 13.08  \\ 
J2157$^{* \dagger}$ & $0.041^{+0.001}_{-0.001}$ & $0.038^{+0.001}_{-0.001}$ & $1.235^{+0.014}_{-0.014}$ & $4.54^{+0.13}_{-0.14}$ & $4.21^{+0.10}_{-0.11}$ & $137.27^{+1.56}_{-1.52}$ & 1.07  \\ 
J2321$^{*}$ & $1.002^{+0.027}_{-0.038}$ & $0.987^{+0.027}_{-0.037}$ & $0.998^{+0.026}_{-0.037}$ & $102.76^{+2.75}_{-3.86}$ & $101.23^{+2.74}_{-3.78}$ & $102.33^{+2.70}_{-3.79}$ & 2.17  \\ 
\hline
\end{tabular}
\end{center}
\endgroup
\end{table*}

\begin{table*}
\begingroup
\renewcommand{\arraystretch}{1.25}
\begin{center}
\caption{Modeled absolute-value lensing magnifications ($\mu$) and apparent magnitudes ($m$) for each lensed quasar image.  For the identifications of each image for each system, see Figures~\ref{fig:model_figure0}, \ref{fig:model_figure1}, and \ref{fig:model_figure2}.}\label{tab:mags}
\begin{tabular}{c|cccc:cccc}
\hline
Name & $\mu_{\rm A}$ & $\mu_{\rm B}$ & $\mu_{\rm C}$ & $\mu_{\rm D}$ & $m_{\rm A}$ & $m_{\rm B}$ & $m_{\rm C}$ & $m_{\rm D}$ \\
 & & & & & [mag] & [mag] & [mag] & [mag]\\
\hline
J0316 & $27.52^{+1.07}_{-0.98}$ & $11.95^{+0.61}_{-0.47}$ & $19.24^{+0.69}_{-0.62}$ &  & $20.17^{+0.02}_{-0.02}$ & $20.84^{+0.03}_{-0.03}$ & $21.20^{+0.03}_{-0.03}$ &   \\ 
J0429 & $11.39^{+0.92}_{-0.68}$ & $13.59^{+1.12}_{-0.83}$ & $10.57^{+0.79}_{-0.59}$ & $4.85^{+0.47}_{-0.35}$ & $18.00^{+0.01}_{-0.01}$ & $18.44^{+0.01}_{-0.01}$ & $18.45^{+0.01}_{-0.01}$ & $19.31^{+0.01}_{-0.01}$  \\ 
J0457 & $2.56^{+0.04}_{-0.04}$ & $3.58^{+0.07}_{-0.07}$ & $3.45^{+0.05}_{-0.05}$ &  & $18.46^{+0.01}_{-0.01}$ & $18.04^{+0.01}_{-0.01}$ & $18.38^{+0.01}_{-0.01}$ &   \\ 
J0607 & $3.62^{+0.26}_{-0.26}$ & $11.18^{+0.93}_{-0.80}$ & $4.55^{+0.44}_{-0.36}$ & $3.02^{+0.15}_{-0.16}$ & $22.56^{+0.23}_{-0.18}$ & $20.43^{+0.05}_{-0.05}$ & $23.00^{+0.51}_{-0.34}$ & $21.94^{+0.16}_{-0.13}$  \\ 
J0608 & $7.29^{+0.52}_{-0.36}$ & $4.38^{+0.49}_{-0.32}$ & $2.52^{+0.18}_{-0.11}$ & $2.45^{+0.19}_{-0.16}$ & $17.97^{+0.01}_{-0.01}$ & $18.42^{+0.01}_{-0.01}$ & $18.93^{+0.01}_{-0.01}$ & $19.05^{+0.01}_{-0.01}$  \\ 
J0719 & $4.69^{+0.20}_{-0.22}$ & $7.05^{+0.30}_{-0.33}$ & $8.97^{+0.35}_{-0.39}$ & $1.48^{+0.26}_{-0.25}$ & $19.70^{+0.01}_{-0.01}$ & $19.28^{+0.01}_{-0.01}$ & $18.96^{+0.01}_{-0.01}$ & $21.81^{+0.10}_{-0.09}$  \\ 
J0722 & $5.69^{+0.54}_{-0.49}$ & $15.10^{+1.49}_{-1.31}$ & $19.43^{+1.77}_{-1.58}$ & $5.86^{+0.66}_{-0.60}$ & $19.55^{+0.01}_{-0.01}$ & $18.19^{+0.01}_{-0.01}$ & $17.80^{+0.01}_{-0.01}$ & $19.51^{+0.02}_{-0.02}$  \\ 
J0756 & $10.06^{+0.98}_{-0.93}$ & $10.60^{+1.12}_{-0.98}$ & $4.40^{+0.41}_{-0.39}$ & $1.95^{+0.24}_{-0.19}$ & $19.07^{+0.01}_{-0.01}$ & $19.94^{+0.05}_{-0.04}$ & $20.39^{+0.04}_{-0.03}$ & $22.02^{+0.08}_{-0.08}$  \\ 
J0803 & $2.24^{+0.07}_{-0.06}$ & $4.87^{+0.19}_{-0.32}$ & $4.91^{+0.17}_{-0.28}$ & $0.79^{+0.03}_{-0.02}$ & $18.42^{+0.03}_{-0.03}$ & $17.84^{+0.02}_{-0.03}$ & $18.07^{+0.03}_{-0.02}$ & $19.07^{+0.04}_{-0.04}$  \\ 
J0833 & $5.40^{+0.53}_{-0.45}$ & $8.35^{+1.01}_{-0.79}$ & $6.38^{+0.73}_{-0.59}$ & $0.95^{+0.13}_{-0.11}$ & $21.68^{+0.03}_{-0.03}$ & $21.19^{+0.04}_{-0.04}$ & $20.86^{+0.03}_{-0.02}$ & $23.15^{+0.11}_{-0.08}$  \\ 
J1258 & $13.21^{+0.67}_{-0.68}$ & $18.13^{+0.91}_{-0.87}$ & $11.02^{+0.58}_{-0.45}$ & $2.54^{+0.15}_{-0.14}$ & $22.85^{+0.32}_{-0.20}$ & $24.52^{+1.52}_{-1.12}$ & $21.89^{+0.20}_{-0.15}$ & $22.54^{+0.16}_{-0.17}$  \\ 
J1640 & $19.81^{+0.24}_{-0.23}$ & $23.75^{+0.28}_{-0.28}$ & $4.07^{+0.03}_{-0.03}$ & $2.09^{+0.01}_{-0.01}$ & $20.44^{+0.03}_{-0.03}$ & $19.72^{+0.01}_{-0.02}$ & $27.29^{+1.23}_{-1.04}$ & $22.45^{+0.10}_{-0.09}$  \\ 
J1651 & $28.63^{+2.39}_{-1.98}$ & $28.89^{+2.18}_{-1.81}$ & $33.08^{+2.71}_{-2.25}$ & $42.10^{+3.27}_{-2.70}$ & $20.11^{+0.01}_{-0.01}$ & $19.56^{+0.01}_{-0.01}$ & $19.64^{+0.01}_{-0.01}$ & $18.75^{+0.01}_{-0.01}$  \\ 
J2017 & $1.23^{+0.35}_{-0.24}$ & $2.63^{+0.62}_{-0.43}$ & $1.92^{+0.46}_{-0.32}$ & $1.71^{+0.33}_{-0.23}$ & $20.22^{+0.30}_{-0.32}$ & $18.77^{+0.09}_{-0.07}$ & $18.78^{+0.05}_{-0.05}$ & $20.95^{+1.10}_{-0.67}$  \\ 
J2103 & $2.62^{+0.27}_{-0.17}$ & $7.88^{+0.96}_{-0.62}$ & $8.53^{+0.95}_{-0.66}$ & $1.33^{+0.23}_{-0.14}$ & $20.04^{+0.02}_{-0.01}$ & $20.89^{+0.09}_{-0.08}$ & $18.62^{+0.01}_{-0.01}$ & $20.71^{+0.03}_{-0.03}$  \\ 
J2157 & $4.83^{+0.11}_{-0.10}$ & $12.10^{+0.36}_{-0.31}$ & $11.12^{+0.28}_{-0.26}$ & $0.92^{+0.02}_{-0.02}$ & $18.33^{+0.01}_{-0.01}$ & $19.10^{+0.02}_{-0.02}$ & $17.64^{+0.01}_{-0.01}$ & $20.82^{+0.03}_{-0.03}$  \\ 
J2321 & $2.01^{+0.11}_{-0.08}$ & $4.70^{+0.44}_{-0.30}$ & $9.45^{+0.71}_{-0.54}$ & $4.51^{+0.32}_{-0.29}$ & $22.52^{+0.40}_{-0.34}$ & $20.71^{+0.08}_{-0.06}$ & $20.95^{+0.10}_{-0.13}$ & $21.57^{+0.12}_{-0.08}$  \\ 
\hline
\end{tabular}
\end{center}
\endgroup
\end{table*}

\section{Conclusions} \label{sec:conclusion}
In this paper, we have presented the uniform gravitational lens modeling of 17 recently discovered lensed quasar systems (2 triply-imaged and 15 quadruply-imaged) observed in the near-infrared \textit{HST} WFC3/IR F160W band (program PID: 17916, PI: T. Treu).  We compiled all prior information for these systems and performed a uniform pipeline modeling scheme (built around \textsc{Lenstronomy}).  We constrain the mass and light profiles of the primary (and secondary, if applicable) lens galaxies of each system, leveraging the superior arc information of the IR imaging.  Despite the vastly different configurations of the sample of lensed quasars, our pipeline is able to successfully construct preliminary models for each system, with systematic uncertainties estimated by probing different shapelet hyper-parameterizations.

From our models, we are able to predict their time delays, neglecting the effects of the MSD, which can only be resolved through additional kinematic observations.  We estimate the combined time-delay distance uncertainties based on Fermat potential uncertainty and time delay measurement precision for an hypothetical follow-up monitoring campaign. We identify six systems that have excellent potential for time-delay cosmography (with expected uncertainties of $\leq 3\%$), and five with good potential (between $3\%$ and $7\%$).  Since the time-delay distance uncertainty directly contributes to the $H_0$ uncertainty, our analysis identifies which systems have the greatest potential to constrain $H_0$.  To this end, we recommend prioritizing follow-up campaigns on the six excellent systems (J0316-4106, J0719+5255, SDSSJ1640+1932, GRALJ1651-0417, DECALSJ2157-4201, and DESIJ2321-0330), as well as the five good systems (J0457-7820, J0608+4229, J0803+3908, J0833+2612, and DELVEJ1258-0319).

In order to perform an unbiased measurement of $H_0$ from a lensed quasar system, one must obtain high-resolution imaging to construct the mass model, IFU spectroscopy to constrain the MSD, and time-delay measurements via high-cadence monitoring programs.  While in this paper we obtain the preliminary mass models for 17 such systems, we also narrow the candidates for future follow-up programs to obtain the necessary data to constrain $H_0$.  The sample of 11 systems recommended for future cosmography studies has the potential to contribute to a precise and accurate late-Universe measurement of $H_0$, capable of resolving the Hubble tension.

    

\begin{acknowledgements}
We thank all the friends of the TDCOSMO collaboration for useful feedback that improved this manuscript.
This research is based on observations made with the NASA/ESA Hubble Space Telescope obtained from the Space Telescope Science Institute, which is operated by the Association of Universities for Research in Astronomy, Inc., under NASA contract NAS 5–26555. These observations are associated with program(s) HST-GO-17916.  Support for US investigators in program \#17916 was provided by NASA through a grant from the Space Telescope Science Institute, which is operated by the Association of Universities for Research in Astronomy, Inc., under NASA contract NAS 5-26555.
KCW is supported by JSPS KAKENHI Grant Numbers JP24H00221, JP24K07089.

\end{acknowledgements}

\appendix
\nolinenumbers
\section{Additional model figures and tables}\label{app:A}
\begin{figure*}
\begin{center}
 \includegraphics[width=1\linewidth]{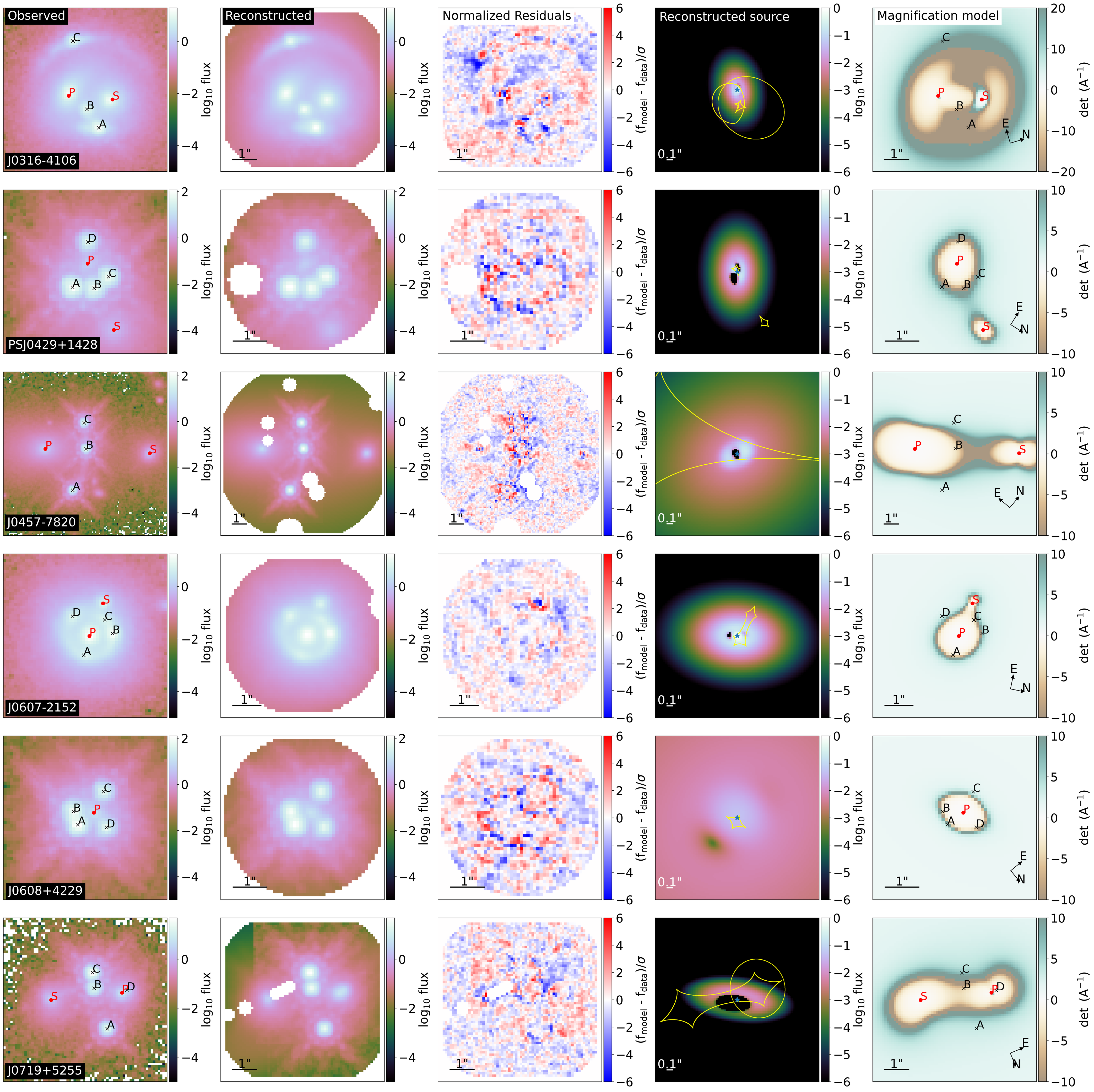}
 \caption{The best-fitting lens models for the first six (of 17) lensed quasars in our sample, using the \textit{HST} F160W IR filter.  The first column shows the \textit{HST} F160W observed image. The second column shows the reconstructed image from our best-fitting model. The third column shows the normalized residuals between the first two plots. The fourth column shows the zoomed-in reconstructed source.  Lastly, the fifth column shows the magnification map of the lensing profile.  The lensed quasar images (A, B, C, D), the primary lens (P), and the secondary lens (S) are labeled on the rightmost and leftmost panels. The remaining systems are shown in Figures~\ref{fig:model_figure1} and \ref{fig:model_figure2}.}
 \label{fig:model_figure0}
\end{center}
\end{figure*}

\begin{figure*}
\begin{center}
 \includegraphics[width=1\linewidth]{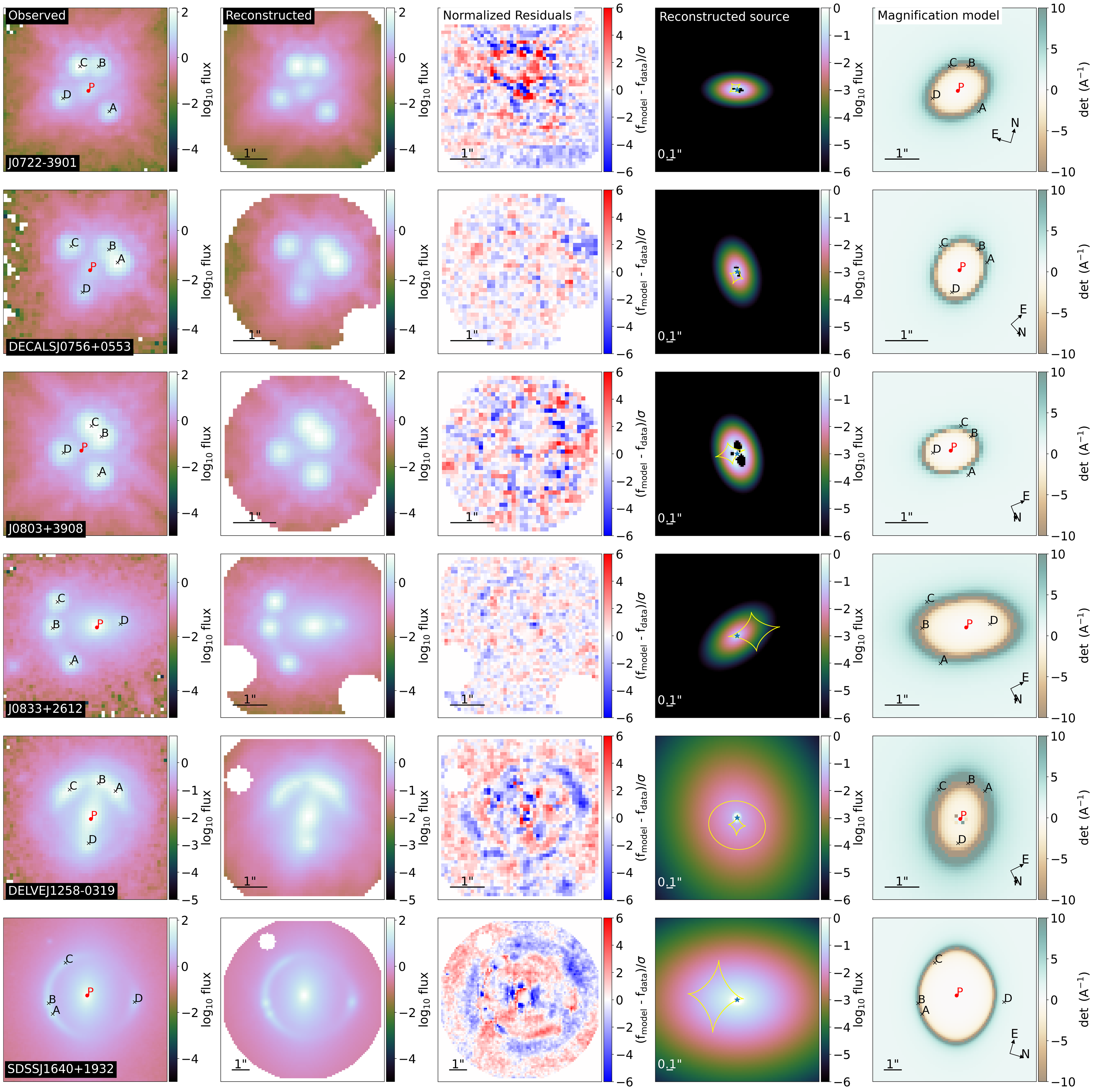}
 \caption{The best-fitting lens models for the next six (of 17) lensed quasars in our sample, using the \textit{HST} F160W IR filter.  See the caption of Figure~\ref{fig:model_figure0} for the full description.  }
 \label{fig:model_figure1}
\end{center}
\end{figure*}

\begin{figure*}
\begin{center}
 \includegraphics[width=1\linewidth]{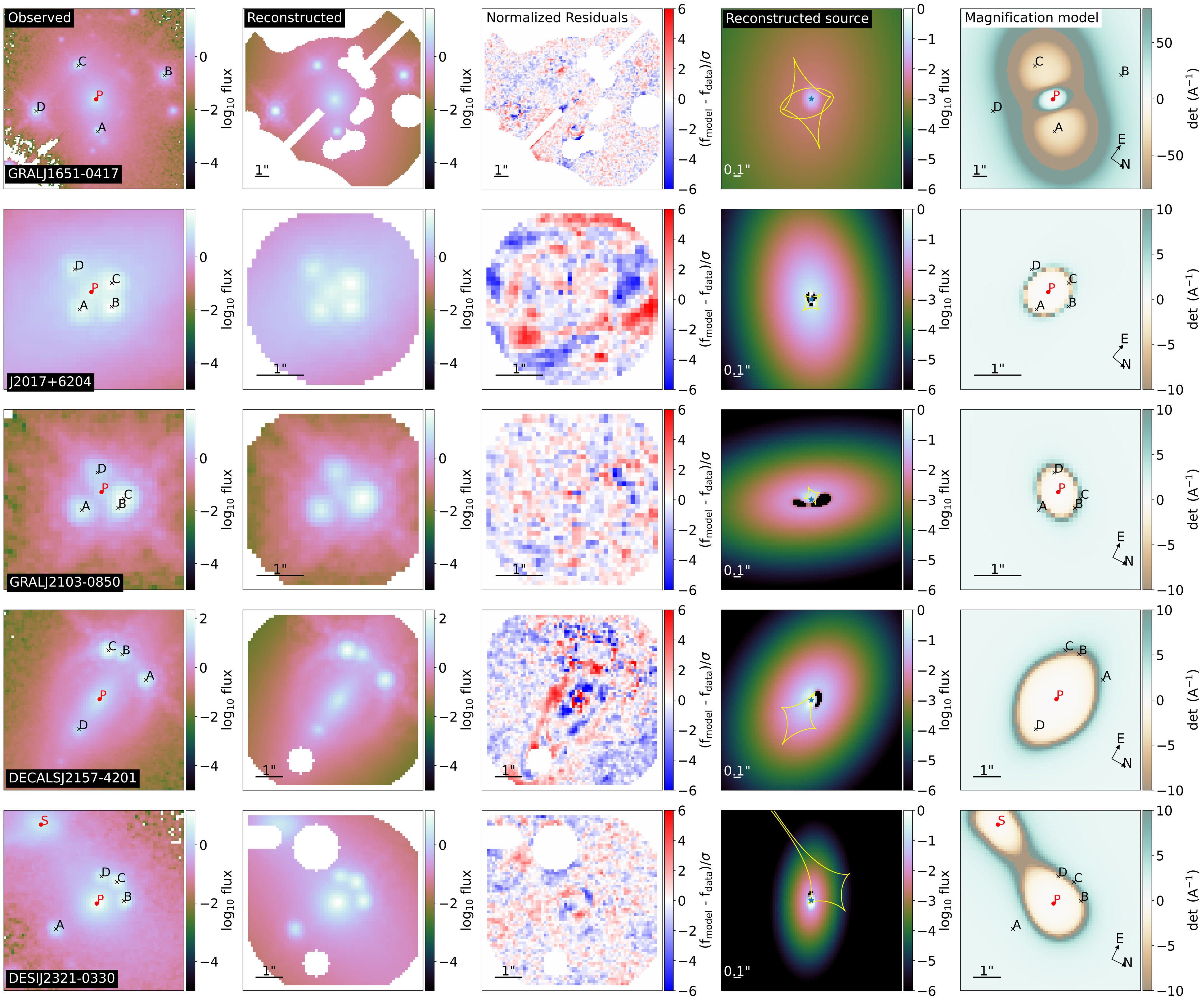}
 \caption{The best-fitting lens models for the last five (of 17) lensed quasars in our sample, using the \textit{HST} F160W IR filter.  See the caption of Figure~\ref{fig:model_figure0} for the full description.  }
 \label{fig:model_figure2}
\end{center}
\end{figure*}

\begin{figure*}
\begin{center}
 \includegraphics[width=1\linewidth]{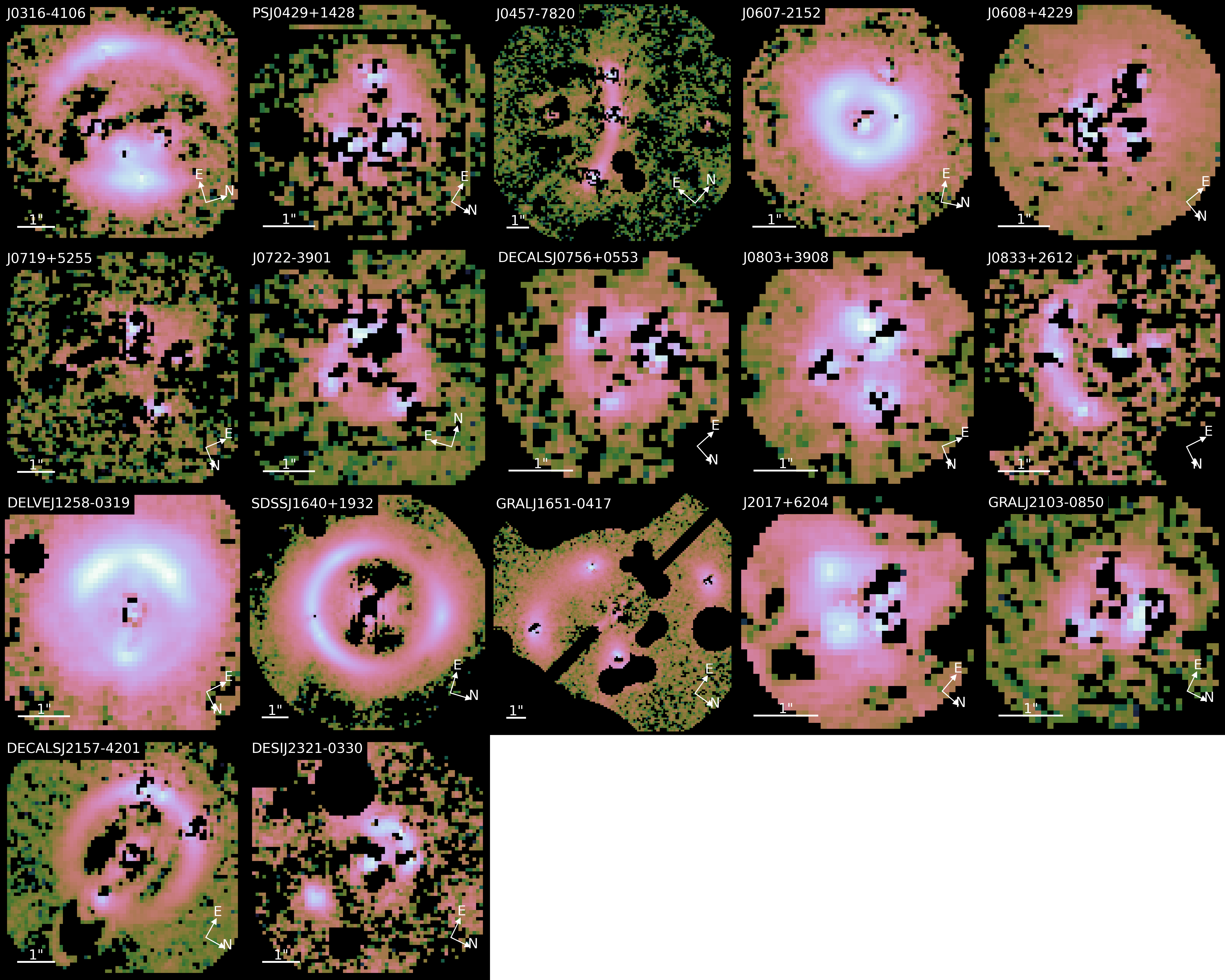}
 \caption{The extended emission of the lensed source galaxy for each system of our sample.  Here we subtract the lens light profile and the PSF images from the data,  isolating only the source light arcs.  These extended emissions visually illustrates the lensing information within each of our lensed quasar system.}
 \label{fig:college_source}
\end{center}
\end{figure*}

In this section, we provide the best-fitting lens models for our sample of 17 lensed quasars in Figures~\ref{fig:model_figure0}, \ref{fig:model_figure1}, and \ref{fig:model_figure2}.  In Figure~\ref{fig:college_source}, we present the data image with the modeled lens light and point sources removed, revealing only the source galaxy lensed emission.  We also provide the full lens and light model posteriors in Tables~\ref{tab:primary_bulge},\ref{tab:primary_halo}, \ref{tab:secondary_bulge}, and \ref{tab:secondary_halo}.  Finally, we provide the convergence and shear values for each image position, useful for future microlensing studies, in Table~\ref{tab:kappa_gamma}.

\begin{table*}
\begingroup
\renewcommand{\arraystretch}{1.25}
\begin{center}
\caption{Modeling results for the S\'ersic profile of the primary-lens bulge light component.  $A_{\rm b, p}$ is the surface brightness, $R_{\rm b, p}$ is the effective radius intrinsic to the S\'ersic profile, $n_{\rm b, p}$ is the S\'ersic index, $q_{\rm b, p}$ is the semi-major to semi-minor axis ratio, $\phi_{\rm b, p}$ is the East-of-North position angle, and $m_{\rm e}$ is the apparent magnitude of the additional lens light component (where applicable; see Section~\ref{subsec:sys_exemptions}).}\label{tab:primary_bulge}
\begin{tabular}{c|cccccc}
\hline
Name & $A_{\rm b, p}$ & $R_{\rm b, p}$ & $n_{\rm b, p}$ & $q_{\rm b, p}$ & $\phi_{\rm b, p}$ & $m_{\rm e}$ \\
 & [mag] [$''$]$^{-2}$ & [$''$] & & & [$^\circ$] & [mag] \\
\hline
J0316 & $14.00^{+0.10}_{-0.11}$ & $0.149^{+0.008}_{-0.007}$ & $3.20^{+0.20}_{-0.15}$ & $0.590^{+0.002}_{-0.002}$ & $33.9^{+0.7}_{-0.9}$ &   \\ 
J0429 & $15.46^{+0.08}_{-0.08}$ & $0.237^{+0.010}_{-0.010}$ & $2.27^{+0.14}_{-0.13}$ & $0.630^{+0.013}_{-0.012}$ & $-2.6^{+2.1}_{-2.1}$ &   \\ 
J0457 & $15.22^{+0.03}_{-0.03}$ & $0.362^{+0.007}_{-0.007}$ & $1.93^{+0.03}_{-0.03}$ & $0.831^{+0.006}_{-0.006}$ & $-56.3^{+1.2}_{-1.2}$ &   \\ 
J0607 & $14.57^{+0.17}_{-0.20}$ & $0.304^{+0.025}_{-0.027}$ & $3.08^{+0.42}_{-0.36}$ & $0.823^{+0.014}_{-0.013}$ & $38.6^{+2.3}_{-2.3}$ & $21.31^{+0.44}_{-0.30}$  \\ 
J0608 & $14.18^{+0.08}_{-0.07}$ & $0.212^{+0.009}_{-0.008}$ & $1.60^{+0.09}_{-0.08}$ & $0.479^{+0.011}_{-0.010}$ & $66.1^{+0.6}_{-0.6}$ &   \\ 
J0719 & $14.33^{+0.10}_{-0.10}$ & $0.141^{+0.005}_{-0.005}$ & $2.96^{+0.21}_{-0.19}$ & $0.805^{+0.014}_{-0.014}$ & $26.8^{+3.0}_{-2.8}$ &   \\ 
J0722 & $12.79^{+0.02}_{-0.01}$ & $0.098^{+0.001}_{-0.001}$ & $7.58^{+0.34}_{-0.29}$ & $0.482^{+0.010}_{-0.010}$ & $-40.9^{+0.5}_{-0.5}$ &   \\ 
J0756 & $15.19^{+0.22}_{-0.22}$ & $0.149^{+0.021}_{-0.018}$ & $1.57^{+0.36}_{-0.31}$ & $0.503^{+0.039}_{-0.032}$ & $-25.5^{+3.0}_{-2.4}$ &   \\ 
J0803 & $14.77^{+0.18}_{-0.14}$ & $0.107^{+0.011}_{-0.007}$ & $2.18^{+0.48}_{-0.55}$ & $0.956^{+0.021}_{-0.040}$ & $-62.4^{+7.8}_{-7.7}$ &   \\ 
J0833 & $13.02^{+0.09}_{-0.05}$ & $0.099^{+0.002}_{-0.001}$ & $2.25^{+0.44}_{-0.23}$ & $0.539^{+0.015}_{-0.015}$ & $22.5^{+1.0}_{-1.0}$ & $24.13^{+1.37}_{-1.26}$  \\ 
J1258 & $15.55^{+0.11}_{-0.10}$ & $0.099^{+0.003}_{-0.001}$ & $9.23^{+0.58}_{-1.00}$ & $0.244^{+0.017}_{-0.008}$ & $-155.0^{+1.2}_{-1.3}$ &   \\ 
J1640 & $13.77^{+0.01}_{-0.01}$ & $0.996^{+0.003}_{-0.003}$ & $4.49^{+0.03}_{-0.03}$ & $0.568^{+0.001}_{-0.001}$ & $-192.8^{+0.1}_{-0.1}$ & $25.80^{+1.47}_{-0.88}$  \\ 
J1651 & $15.68^{+0.04}_{-0.04}$ & $0.810^{+0.019}_{-0.018}$ & $2.96^{+0.04}_{-0.04}$ & $0.847^{+0.003}_{-0.003}$ & $-225.1^{+0.5}_{-0.5}$ &   \\ 
J2017 & $13.43^{+0.11}_{-0.11}$ & $0.201^{+0.012}_{-0.012}$ & $2.02^{+0.19}_{-0.16}$ & $0.908^{+0.033}_{-0.034}$ & $-164.7^{+7.4}_{-7.2}$ &   \\ 
J2103 & $13.47^{+0.06}_{-0.06}$ & $0.098^{+0.001}_{-0.001}$ & $0.51^{+0.12}_{-0.09}$ & $0.591^{+0.050}_{-0.055}$ & $127.1^{+3.6}_{-3.8}$ &   \\ 
J2157 & $15.40^{+0.02}_{-0.02}$ & $0.774^{+0.002}_{-0.002}$ & $5.27^{+0.07}_{-0.05}$ & $0.496^{+0.004}_{-0.004}$ & $8.4^{+0.1}_{-0.1}$ &   \\ 
J2321 & $14.75^{+0.03}_{-0.03}$ & $0.349^{+0.005}_{-0.005}$ & $1.67^{+0.07}_{-0.06}$ & $0.999^{+0.001}_{-0.001}$ & $113.5^{+7.6}_{-7.5}$ & $20.93^{+0.08}_{-0.07}$  \\ 
\hline
\end{tabular}
\end{center}
\endgroup
\end{table*}

\begin{table*}
\begingroup
\renewcommand{\arraystretch}{1.25}
\begin{center}
\caption{Modeling results for the S\'ersic profile of the primary-lens halo light component.  $A_{\rm h, p}$ is the surface brightness, $R_{\rm h, p}$ is the effective radius intrinsic to the S\'ersic profile, $n_{\rm h, p}$ is the S\'ersic index, $q_{\rm h, p}$ is the semi-major to semi-minor axis ratio, and $\phi_{\rm h, p}$ is the East-of-North position angle.}\label{tab:primary_halo}
\begin{tabular}{c|ccccc}
\hline
Name & $A_{\rm h, p}$ & $R_{\rm h, p}$ & $n_{\rm h, p}$ & $q_{\rm h, p}$ & $\phi_{\rm h, p}$ \\
 & [mag] [$''$]$^{-2}$ & [$''$] & & & [$^\circ$] \\
\hline
J0316 & $17.00^{+0.06}_{-0.06}$ & $1.077^{+0.026}_{-0.024}$ & $3.05^{+0.11}_{-0.10}$ & $0.589^{+0.002}_{-0.002}$ & $45.9^{+0.3}_{-0.3}$  \\ 
J0429 & $19.14^{+0.04}_{-0.04}$ & $1.501^{+0.075}_{-0.095}$ & $0.19^{+0.02}_{-0.01}$ & $0.219^{+0.020}_{-0.014}$ & $-31.8^{+1.0}_{-1.1}$  \\ 
J0457 & $18.01^{+0.03}_{-0.03}$ & $2.676^{+0.048}_{-0.046}$ & $1.62^{+0.03}_{-0.03}$ & $0.489^{+0.003}_{-0.003}$ & $-38.4^{+0.2}_{-0.2}$  \\ 
J0607 & $16.84^{+0.09}_{-0.09}$ & $1.001^{+0.028}_{-0.027}$ & $0.75^{+0.06}_{-0.06}$ & $0.810^{+0.014}_{-0.012}$ & $127.4^{+1.4}_{-1.4}$  \\ 
J0608 & $17.72^{+0.14}_{-0.13}$ & $0.722^{+0.024}_{-0.037}$ & $0.21^{+0.02}_{-0.02}$ & $0.473^{+0.012}_{-0.012}$ & $72.3^{+1.0}_{-0.9}$  \\ 
J0719 & $20.30^{+0.08}_{-0.07}$ & $2.205^{+0.109}_{-0.102}$ & $0.40^{+0.07}_{-0.05}$ & $0.208^{+0.008}_{-0.006}$ & $-6.6^{+1.0}_{-1.2}$  \\ 
J0722 & $15.95^{+0.04}_{-0.04}$ & $0.458^{+0.007}_{-0.007}$ & $0.34^{+0.02}_{-0.02}$ & $0.291^{+0.008}_{-0.008}$ & $-44.5^{+0.3}_{-0.3}$  \\ 
J0756 & $17.06^{+0.36}_{-0.28}$ & $0.398^{+0.044}_{-0.035}$ & $1.18^{+0.19}_{-0.20}$ & $0.339^{+0.027}_{-0.030}$ & $-33.9^{+0.9}_{-0.9}$  \\ 
J0803 & $16.95^{+0.16}_{-0.16}$ & $0.154^{+0.023}_{-0.025}$ & $0.11^{+0.02}_{-0.01}$ & $0.243^{+0.041}_{-0.027}$ & $-60.7^{+1.4}_{-1.5}$  \\ 
J0833 & $18.14^{+0.13}_{-0.10}$ & $1.265^{+0.067}_{-0.053}$ & $1.93^{+0.20}_{-0.16}$ & $0.507^{+0.008}_{-0.008}$ & $25.5^{+0.5}_{-0.5}$  \\ 
J1258 & $14.19^{+0.01}_{-0.01}$ & $0.251^{+0.002}_{-0.002}$ & $1.27^{+0.03}_{-0.03}$ & $0.234^{+0.003}_{-0.003}$ & $-229.3^{+0.1}_{-0.1}$  \\ 
J1640 & $18.68^{+0.01}_{-0.02}$ & $9.928^{+0.054}_{-0.127}$ & $1.25^{+0.02}_{-0.02}$ & $0.568^{+0.001}_{-0.001}$ & $-270.0^{+0.1}_{-0.1}$  \\ 
J1651 & $18.62^{+0.06}_{-0.05}$ & $4.065^{+0.136}_{-0.124}$ & $1.15^{+0.05}_{-0.04}$ & $0.446^{+0.007}_{-0.008}$ & $-230.0^{+0.2}_{-0.2}$  \\ 
J2017 & $15.68^{+0.01}_{-0.01}$ & $1.501^{+0.012}_{-0.012}$ & $0.50^{+0.01}_{-0.01}$ & $0.408^{+0.005}_{-0.005}$ & $-157.0^{+0.1}_{-0.1}$  \\ 
J2103 & $17.25^{+0.12}_{-0.12}$ & $0.375^{+0.023}_{-0.029}$ & $0.19^{+0.02}_{-0.01}$ & $0.442^{+0.040}_{-0.043}$ & $-48.6^{+1.1}_{-1.2}$  \\ 
J2157 & $15.83^{+0.02}_{-0.03}$ & $0.775^{+0.002}_{-0.002}$ & $1.57^{+0.04}_{-0.04}$ & $0.203^{+0.003}_{-0.002}$ & $5.1^{+0.1}_{-0.1}$  \\ 
J2321 & $18.91^{+0.05}_{-0.04}$ & $4.263^{+0.169}_{-0.136}$ & $0.90^{+0.04}_{-0.04}$ & $0.365^{+0.012}_{-0.013}$ & $112.5^{+0.4}_{-0.4}$  \\ 
\hline
\end{tabular}
\end{center}
\endgroup
\end{table*}


\begin{table*}
\begingroup
\renewcommand{\arraystretch}{1.25}
\begin{center}
\caption{Modeling results for the S\'ersic profile of the secondary-lens bulge light component.  $A_{\rm b, s}$ is the surface brightness, $R_{\rm b, s}$ is the effective radius intrinsic to the S\'ersic profile, $n_{\rm b, s}$ is the S\'ersic index, $q_{\rm b, s}$ is the semi-major to semi-minor axis ratio, and $\phi_{\rm b, s}$ is the East-of-North position angle.}\label{tab:secondary_bulge}
\begin{tabular}{c|ccccc}
\hline
Name & $A_{\rm b, s}$ & $R_{\rm b, s}$ & $n_{\rm b, s}$ & $q_{\rm b, s}$ & $\phi_{\rm b, s}$ \\
 & [mag] [$''$]$^{-2}$ & [$''$] & & & [$^\circ$] \\
\hline
J0316 & $14.76^{+0.09}_{-0.08}$ & $0.234^{+0.010}_{-0.008}$ & $3.80^{+0.27}_{-0.21}$ & $0.729^{+0.006}_{-0.007}$ & $-70.6^{+1.1}_{-1.0}$  \\ 
J0429 & $16.20^{+0.08}_{-0.08}$ & $0.394^{+0.022}_{-0.020}$ & $1.51^{+0.09}_{-0.08}$ & $0.981^{+0.008}_{-0.009}$ & $84.3^{+16.9}_{-25.2}$  \\ 
J0457 & $15.45^{+0.03}_{-0.03}$ & $0.278^{+0.004}_{-0.004}$ & $3.54^{+0.10}_{-0.08}$ & $0.855^{+0.005}_{-0.005}$ & $-97.0^{+1.1}_{-1.1}$  \\ 
J0607 & $14.06^{+0.03}_{-0.02}$ & $0.098^{+0.001}_{-0.001}$ & $5.72^{+1.36}_{-0.85}$ & $0.880^{+0.022}_{-0.023}$ & $32.8^{+4.9}_{-4.3}$  \\ 
J0719 & $15.80^{+0.26}_{-0.29}$ & $0.177^{+0.026}_{-0.024}$ & $8.20^{+1.16}_{-1.19}$ & $0.757^{+0.024}_{-0.022}$ & $-269.6^{+0.7}_{-0.3}$  \\ 
J2321 & $15.53^{+0.02}_{-0.02}$ & $0.377^{+0.005}_{-0.005}$ & $0.75^{+0.02}_{-0.02}$ & $0.675^{+0.007}_{-0.008}$ & $74.0^{+0.9}_{-1.0}$  \\ 
\hline
\end{tabular}
\end{center}
\endgroup
\end{table*}

\begin{table*}
\begingroup
\renewcommand{\arraystretch}{1.25}
\begin{center}
\caption{Modeling results for the S\'ersic profile of the secondary-lens halo light component.  $A_{\rm h, s}$ is the surface brightness, $R_{\rm h, s}$ is the effective radius intrinsic to the S\'ersic profile, $n_{\rm h, s}$ is the S\'ersic index, $q_{\rm h, s}$ is the semi-major to semi-minor axis ratio, and $\phi_{\rm h, s}$ is the East-of-North position angle.}\label{tab:secondary_halo}
\begin{tabular}{c|ccccc}
\hline
Name & $A_{\rm h, s}$ & $R_{\rm h, s}$ & $n_{\rm h, s}$ & $q_{\rm h, s}$ & $\phi_{\rm h, s}$ \\
 & [mag] [$''$]$^{-2}$ & [$''$] & & & [$^\circ$] \\
\hline
J0316 & $17.80^{+0.05}_{-0.04}$ & $1.385^{+0.028}_{-0.028}$ & $1.92^{+0.07}_{-0.06}$ & $0.725^{+0.006}_{-0.008}$ & $1.1^{+1.0}_{-0.9}$  \\ 
J0429 & $17.94^{+0.07}_{-0.06}$ & $1.138^{+0.028}_{-0.035}$ & $0.22^{+0.02}_{-0.02}$ & $0.939^{+0.027}_{-0.030}$ & $91.9^{+14.2}_{-29.2}$  \\ 
J0457 & $21.61^{+0.07}_{-0.07}$ & $9.756^{+0.174}_{-0.267}$ & $0.99^{+0.06}_{-0.05}$ & $0.239^{+0.018}_{-0.017}$ & $-67.3^{+1.3}_{-1.3}$  \\ 
J0607 & $19.37^{+0.06}_{-0.06}$ & $2.837^{+0.377}_{-0.424}$ & $0.18^{+0.01}_{-0.01}$ & $0.244^{+0.069}_{-0.035}$ & $35.9^{+1.6}_{-1.6}$  \\ 
J0719 & $18.10^{+0.06}_{-0.05}$ & $0.582^{+0.014}_{-0.013}$ & $0.34^{+0.05}_{-0.05}$ & $0.699^{+0.021}_{-0.024}$ & $-203.0^{+2.3}_{-2.6}$  \\ 
J2321 & $18.37^{+0.05}_{-0.06}$ & $1.391^{+0.033}_{-0.037}$ & $0.22^{+0.03}_{-0.02}$ & $0.377^{+0.017}_{-0.017}$ & $104.9^{+0.8}_{-0.8}$  \\ 
\hline
\end{tabular}
\end{center}
\endgroup
\end{table*}

\begin{table*}
\begingroup
\renewcommand{\arraystretch}{1.25}
\begin{center}
\caption{Modeled lensing convergence ($\kappa$) and shear ($\gamma_{\rm shear}$) for each lensed quasar image.  For the identifications of each image for each system, see Figures~\ref{fig:model_figure0}, \ref{fig:model_figure1}, and \ref{fig:model_figure2}.}\label{tab:kappa_gamma}
\begin{tabular}{c|cccc:cccc}
\hline
Name & $\kappa_{\rm A}$ & $\kappa_{\rm B}$ & $\kappa_{\rm C}$ & $\kappa_{\rm D}$ & $\gamma_{\rm shear, A}$ & $\gamma_{\rm shear, B}$ & $\gamma_{\rm shear, C}$ & $\gamma_{\rm shear, D}$ \\
\hline
J0316 & $0.907^{+0.002}_{-0.002}$ & $1.067^{+0.002}_{-0.002}$ & $0.621^{+0.005}_{-0.005}$ &  & $0.212^{+0.004}_{-0.004}$ & $0.297^{+0.006}_{-0.007}$ & $0.303^{+0.004}_{-0.004}$ &   \\ 
J0429 & $0.555^{+0.012}_{-0.010}$ & $0.561^{+0.012}_{-0.012}$ & $0.455^{+0.014}_{-0.011}$ & $0.663^{+0.014}_{-0.015}$ & $0.332^{+0.010}_{-0.009}$ & $0.516^{+0.012}_{-0.014}$ & $0.449^{+0.011}_{-0.012}$ & $0.566^{+0.016}_{-0.020}$  \\ 
J0457 & $0.278^{+0.004}_{-0.004}$ & $0.384^{+0.004}_{-0.004}$ & $0.298^{+0.004}_{-0.004}$ &  & $0.361^{+0.002}_{-0.002}$ & $0.811^{+0.006}_{-0.006}$ & $0.452^{+0.002}_{-0.002}$ &   \\ 
J0607 & $0.384^{+0.023}_{-0.026}$ & $0.384^{+0.019}_{-0.023}$ & $0.478^{+0.021}_{-0.024}$ & $0.293^{+0.016}_{-0.019}$ & $0.809^{+0.032}_{-0.026}$ & $0.539^{+0.021}_{-0.018}$ & $0.701^{+0.026}_{-0.022}$ & $0.411^{+0.014}_{-0.013}$  \\ 
J0608 & $0.220^{+0.030}_{-0.023}$ & $0.374^{+0.032}_{-0.029}$ & $0.157^{+0.026}_{-0.018}$ & $0.395^{+0.043}_{-0.033}$ & $0.687^{+0.024}_{-0.032}$ & $0.788^{+0.028}_{-0.034}$ & $0.560^{+0.018}_{-0.022}$ & $0.881^{+0.028}_{-0.034}$  \\ 
J0719 & $0.526^{+0.009}_{-0.011}$ & $0.697^{+0.007}_{-0.009}$ & $0.613^{+0.008}_{-0.010}$ & $1.817^{+0.040}_{-0.041}$ & $0.105^{+0.003}_{-0.003}$ & $0.484^{+0.012}_{-0.010}$ & $0.195^{+0.007}_{-0.005}$ & $1.162^{+0.054}_{-0.046}$  \\ 
J0722 & $0.400^{+0.025}_{-0.027}$ & $0.530^{+0.024}_{-0.027}$ & $0.448^{+0.026}_{-0.028}$ & $0.626^{+0.022}_{-0.024}$ & $0.429^{+0.020}_{-0.019}$ & $0.536^{+0.028}_{-0.027}$ & $0.503^{+0.026}_{-0.025}$ & $0.558^{+0.031}_{-0.029}$  \\ 
J0756 & $0.357^{+0.026}_{-0.042}$ & $0.411^{+0.028}_{-0.046}$ & $0.331^{+0.025}_{-0.041}$ & $0.538^{+0.041}_{-0.053}$ & $0.561^{+0.039}_{-0.024}$ & $0.664^{+0.048}_{-0.029}$ & $0.473^{+0.033}_{-0.022}$ & $0.846^{+0.064}_{-0.044}$  \\ 
J0803 & $0.225^{+0.010}_{-0.010}$ & $0.296^{+0.011}_{-0.012}$ & $0.266^{+0.011}_{-0.012}$ & $0.456^{+0.018}_{-0.021}$ & $0.395^{+0.006}_{-0.006}$ & $0.838^{+0.016}_{-0.013}$ & $0.577^{+0.009}_{-0.010}$ & $1.248^{+0.019}_{-0.026}$  \\ 
J0833 & $0.376^{+0.027}_{-0.026}$ & $0.539^{+0.023}_{-0.023}$ & $0.445^{+0.025}_{-0.025}$ & $0.967^{+0.032}_{-0.024}$ & $0.452^{+0.018}_{-0.019}$ & $0.576^{+0.029}_{-0.028}$ & $0.389^{+0.017}_{-0.015}$ & $1.030^{+0.064}_{-0.066}$  \\ 
J1258 & $0.542^{+0.011}_{-0.012}$ & $0.617^{+0.011}_{-0.013}$ & $0.498^{+0.011}_{-0.012}$ & $0.852^{+0.012}_{-0.016}$ & $0.366^{+0.011}_{-0.009}$ & $0.449^{+0.014}_{-0.012}$ & $0.401^{+0.010}_{-0.009}$ & $0.645^{+0.020}_{-0.020}$  \\ 
J1640 & $0.149^{+0.002}_{-0.002}$ & $0.134^{+0.002}_{-0.002}$ & $0.217^{+0.003}_{-0.003}$ & $0.096^{+0.001}_{-0.001}$ & $0.880^{+0.002}_{-0.002}$ & $0.841^{+0.002}_{-0.002}$ & $0.927^{+0.002}_{-0.002}$ & $0.581^{+0.002}_{-0.001}$  \\ 
J1651 & $0.996^{+0.001}_{-0.001}$ & $0.810^{+0.007}_{-0.006}$ & $0.957^{+0.002}_{-0.002}$ & $0.841^{+0.006}_{-0.005}$ & $0.187^{+0.007}_{-0.007}$ & $0.036^{+0.001}_{-0.001}$ & $0.179^{+0.007}_{-0.007}$ & $0.040^{+0.001}_{-0.001}$  \\ 
J2017 & $0.167^{+0.116}_{-0.119}$ & $0.120^{+0.091}_{-0.087}$ & $0.152^{+0.108}_{-0.109}$ & $0.099^{+0.076}_{-0.072}$ & $1.227^{+0.158}_{-0.152}$ & $0.630^{+0.060}_{-0.070}$ & $1.113^{+0.128}_{-0.129}$ & $0.483^{+0.034}_{-0.046}$  \\ 
J2103 & $0.232^{+0.026}_{-0.021}$ & $0.344^{+0.023}_{-0.029}$ & $0.271^{+0.025}_{-0.025}$ & $0.534^{+0.035}_{-0.042}$ & $0.456^{+0.022}_{-0.021}$ & $0.748^{+0.027}_{-0.029}$ & $0.643^{+0.026}_{-0.021}$ & $0.986^{+0.050}_{-0.062}$  \\ 
J2157 & $0.276^{+0.008}_{-0.007}$ & $0.393^{+0.007}_{-0.007}$ & $0.307^{+0.008}_{-0.007}$ & $0.652^{+0.011}_{-0.011}$ & $0.564^{+0.006}_{-0.006}$ & $0.672^{+0.007}_{-0.008}$ & $0.625^{+0.006}_{-0.007}$ & $1.098^{+0.013}_{-0.014}$  \\ 
J2321 & $0.268^{+0.018}_{-0.014}$ & $0.455^{+0.024}_{-0.022}$ & $0.435^{+0.023}_{-0.020}$ & $0.491^{+0.025}_{-0.022}$ & $0.192^{+0.004}_{-0.005}$ & $0.714^{+0.025}_{-0.031}$ & $0.462^{+0.019}_{-0.020}$ & $0.695^{+0.024}_{-0.029}$  \\ 
\hline
\end{tabular}
\end{center}
\endgroup
\end{table*}

\bibliographystyle{aa}
\bibliography{biblio}

\end{document}